\documentclass[12pt]{iopart}

\usepackage{graphicx}% Include figure files
\usepackage{dcolumn}% Align table columns on decimal point
\usepackage{bm}% bold math

\begin{document}

\title[Density of states and extent of wave function ... ]{Density of states and extent of wave function: \\ two crucial factors for small polaron hopping conductivity in 1D}

\author{Margarita Dimakogianni, Constantinos Simserides, and Georgios P. Triberis}
\address{University of Athens, Physics Department, Panepistimiopolis, 15784 Zografos, Athens, Greece}
\ead{mdimak@phys.uoa.gr, csimseri@phys.uoa.gr, gtriber@phys.uoa.gr}

\begin{abstract}
We introduce a theoretical model to scrutinize the conductivity of
small polarons in one-dimensional disordered systems, focusing on
two crucial --as will be demonstrated-- factors: the \emph{density
of states} and the \emph{spatial extent of the electronic wave
function}. The investigation is performed for any temperature up
to 300 K and under electric field of arbitrary strength up to the
polaron dissociation limit. To accomplish this task we combine
analytical work with numerical calculations.
\end{abstract}

% 71.38.-k    Polarons and electron-phonon interactions (see also 63.20.K- Phonon interactions in lattice dynamics)
% 72.80.-r    Conductivity of specific materials (for conductivity of metals and alloys, see 72.15.-v)
% 72.80.Ng    Disordered solids
% 73.63.-b    Electronic transport in nanoscale materials and structures (see also 73.23.-b Electronic transport in mesoscopic systems)
% 71.20.-b    Electron density of states and band structure of crystalline solids
% 73.20.At    Surface states, band structure, electron density of states

\maketitle

\section{\label{Sec:intro} Introduction}
The \emph{density of states} (DOS) is the heart of any physical
system in the sense that its structure and magnitude crucially
affect all physical properties. Particularly, this holds for the
response of charge carriers to external stimuli such as electric
or magnetic fields, temperature gradients or temperature
variations, i.e. the transport properties. In disordered
materials, the random distribution of their constituents
drastically affects the character of the carriers and the
transport mechanisms. Under certain circumstances, the presence of
disorder induces carrier localization and hopping becomes the
chief transport mechanism. Hence, the \emph{electronic wave
function spatial extent} ($\alpha^{-1}$), being a measure of the
carrier localization, becomes a parameter of vital importance.

One-dimensional (1D) systems have been recently considered to be
among the most promising materials for nanotechnology. In
particular, an increasing amount of experimental and theoretical
work has been devoted to the electrical properties of 1D amorphous
semiconductors, amorphous carbon, doped polymers, conjugated
polymers and organic materials
\cite{Gleve:1995,Nebel:1992,Godet:2003,Cumings:2004,Tang:2000,Campbell:1999,Mozer:2005,Novikov:1998,Aleshin:2004,Yu:2000,Yu:2001},
\cite[and references therein]{TDRev:2009}.

Given that DNA has been placed among the most promising organic
materials for nanotechnology, Triberis~\emph{et
al.}~\cite{TSK:2005}, studied DNA as a 1D disordered molecular
``wire'' in which small polarons are the charge carriers. Based on
the Generalized Molecular Crystal Model (GMCM)~\cite{TF:1981} and
theoretical percolation arguments, they studied small polaron
hopping along the DNA double helix and in the presence of low
electric field ($F$). Ignoring the effect of correlations, an
analytical expression for the strong temperature ($T$) dependence
of the electrical conductivity ($\sigma$) was obtained which
reproduced the experimental data reported for $\lambda$-DNA
\cite{Tran:2000} and for poly(dA)-poly(dT) DNA~\cite{Yoo:2001} at
high temperatures. The theoretical analysis also permitted the
evaluation of the maximum hopping distance and its $T$-dependence,
supporting the idea of multi-phonon assisted hopping of small
polarons between next nearest neighbors of the DNA molecular
``wire''. Taking into account the effect of correlations ($cr$),
Triberis and Dimakogianni~\cite{TDRev:2009,TD:2009} showed that
$ln \sigma^{cr} \propto T^{-1/2}$ holds for high as well as for
low temperatures. This reproduced the strong $\sigma(T)$ at high
temperatures reported for
$\lambda$-DNA~\cite{Tran:2000,Inomata:2006} and poly(dA)-poly(dT)
DNA~\cite{Yoo:2001}, while, including correlations, the evaluation
of the maximum hopping distance led to systematically longer
values than those evaluated ignoring correlations~\cite{TSK:2005},
supporting experimental evidence for long range charge migration
along the DNA double helix
\cite{Schuster:2000,Carell:2003,Takada:2004}.

In addition, even under moderate electric fields,
strong nonlinearities of $\sigma(F)$
in 1D disordered systems have been observed.
In the variable range hopping regime
and at low temperatures, Fogler and Kelley~\cite{FoglerKelley:2005}
investigated theoretically the effect of a finite electric field
on the resistivity. They took into account the
existence of highly resistive segments (breaks) on the conducting
path of the carriers in 1D systems and found that the role of
the breaks diminishes and eventually becomes insignificant as $F$
increases. Ma \emph{et al.}~\cite{Ma:2007}
described hopping transport and the conductivity of 1D systems with off-diagonal disorder.
Investigating the $T$-dependence of the hopping conductivity, they
showed that it increases with the increase of $T$ taking much
larger values than in the case of the Anderson model with pure
diagonal disorder. They also studied the $F$-dependence of the
conductivity to find that at low $F$ the hopping conductivity
conforms with the ohmic law, but at strong fields it presents
non-ohmic characteristics.

Triberis and Dimakogianni~\cite{TD:2009:F} studied the $\sigma(T,F)$ behaviour
under the influence of moderate electric fields up to $\sim$ 10$^5$ Vm$^{-1}$,
when small polarons are transported in a disordered 1D environment,
at high and low temperatures. The analytical expressions obtained for
$\sigma(F,T)$, were applied to experimental findings concerning
charge transport in polydiacetylene quasi-1D single crystals
\cite{Aleshin:2004}. It was shown that at low electric fields the
hopping conductivity conforms with the ohmic law while increasing
the electric field the conductivity presents non-ohmic
characteristics. The transition from the ohmic to the non-ohmic
behaviour starts for smaller values of $F$ at lower temperatures
and the rate of the increase of $\sigma$ is greater the lower $T$
is. These conclusions were in a qualitative agreement with
theoretical results referred to variable range hopping
\cite{FoglerKelley:2005,Ma:2007,RaikhRuzin:1989}.
Dimakogianni and Triberis~\cite{DT:2010:Fcr} also investigated the
effect of correlations on the non-ohmic behaviour of the small
polaron hopping conductivity in 1D and 3D disordered systems. They
concluded that the inclusion of correlations results to a much
stronger dependence of the conductivity on the magnitude of the
applied electric field compared to the uncorrelated case. The
deviation of the conductivity from the ohmic behaviour appears
twice as fast when correlation effects are taken into account,
for a given applied electric field as the temperature increases.

In the present work, taking into account the directionality
imposed by the electric field on the transport path of the
carriers, we examine the role of the magnitude of the density of
states and the extent of the electronic wave function and
calculate $\sigma$. The aim of the present work, is to investigate
$\sigma(T,F)$ for all reasonable $T$ and $F$ values, i.e. from 10
up to 300 K and up to the $F$ values where polarons cease to
exist. This is done varying the density of states by orders of
magnitude around values which are relevant to common 1D
systems~\cite{HKS:2010,HKS:2011:erratum,Bourbie:2007,Bourbie:2011}
and varying the extent of the electronic wave function from 1 to 5
$\AA$, i.e. reasonable values for common organic
molecules~\cite{HKS:2009,HSK:2009}. We demonstrate that $F$ plays
both a \emph{constructive} energetic role by offering energy for
the carrier jumps and simultaneously a \emph{destructive} role, in
the sense that the stronger it is the more it forces the polaron
to jump opposite to the $\bm F$ direction prohibiting forward
jumps to neighboring sites.

In Section~\ref{Sec:theory} we present our theoretical model
including the basic analytical expressions at \emph{high} and
\emph{low temperatures}. According to the mathematical analysis of
the Generalized Molecular Crystal Model ~\cite{TF:1981,TF:1986},
it is the condition $\hbar\omega_{0}\ll k_{B}T$
($\hbar\omega_{0}\gg k_{B}T$) that determines the \emph{high}
(\emph{low}) \emph{temperature} regime. This mathematical analysis
leads to the evaluation of the intrinsic transition rate, which
differs at high temperatures (multi-phonon assisted hopping),
compared to that at low temperatures (few-phonon assisted
hopping). Which temperature range in real systems is indeed
\emph{high} or \emph{low} depends on the system under study. Our
numerical results together with the relevant discussion are staged
in Section~\ref{Sec:resdis}. In this way we examine the
conductivity of small polarons in one-dimensional disordered
systems, and demonstrate that the \emph{density of states} and the
\emph{spatial extent of the electronic wave function} are two
crucial factors for its behaviour. The temperature and the
electric field ranges that we consider are very broad. In
particular, the electric field is varied from very low up to the
polaron dissociation limit ($\sim 1 \times 10^8$ Vm$^{-1}$).
Finally, in Section~\ref{Sec:conclusion} we state our conclusions.

\section{\label{Sec:theory} Theory}

\subsection{\label{Subsec:GMCM} Generalized Molecular Crystal Model}

In the context of GMCM we consider a 1D deformable ``wire''
consisting of ``molecular lattice sites'' across which small
polarons are transported in the presence of disorder. By
$\epsilon_i (0)$, and $\epsilon_j (0)$ we denote the energies of
an electron on site at vector positions $\mathbf{r}_i$ and
$\mathbf{r}_j$, respectively, if the ``molecular lattice sites''
are constrained not to be displaced in response to the presence of
the electron. Due to the disorder these local electronic energies,
$\epsilon_i (0)$, and $\epsilon_j (0)$ are not equal. The
energetic non-equivalence of the two sites will affect the small
polaron's binding energy, $E_b(i)$, in the sense that, the lower
the local electronic energy is the more localized the electronic
wave function will tend to be and consequently the larger its
binding energy will be. Assuming that the stiffness of the
``molecular lattice'' is unaltered, the difference in binding
energy means a difference in the electron-lattice interaction
parameters $A_i$ and $A_j$ i.e. $E_{i}(\mathbf{x}_i) =
\epsilon_{i}(0) -A_i\mathbf{x}_i$ and $E_{j}(\mathbf{x}_j) =
\epsilon_{j}(0) -A_j\mathbf{x}_j$ with $A_i \neq A_j$. Here,
$E_i(\mathbf{x}_i)$ is the electronic energy of the system of the
electron and the isolated molecule with configurational coordinate
$\mathbf{x}_i$, which represents the deviation of the atoms of the
molecule at position $\mathbf{r}_i$ from their equilibrium
configuration i.e. the local vibrational displacement coordinate.

The GMCM ~\cite{TF:1981} is based on a generalized  ``hopping
model'' Hamiltonian of the form
\begin{equation}\label{GMCM}
< m|H|n >=<m|H_0+V|n>=E_{i,\{n_{\mathbf{k}}\}} \delta_{ij} \delta_
{\{n_{\mathbf{k}}\},\{n_{\mathbf{k}'}\}} + <m|V|n> ,
\end{equation}
the $<m|V|n>$ term ~\cite{TF:1981} is the overlap part of the
Hamiltonian, $|n>=|i,\{n_{\mathbf{k}}\}>$ are the eigenstates of
$H$, and $H_0$ is the zeroth-order (i.e. for electronic overlap
integral of the tight-binding theory $J$=0) Hamiltonian with
corresponding eigenvalues
\begin{equation}\label{energy-1}
E_{i,\{n_{\mathbf{k}}\}}=\epsilon_i (0) -E_b(i) +
\sum_{\mathbf{k}} \hbar\omega_{\mathbf{k}} (n_{\mathbf{k}}
+\frac{1}{2}).
\end{equation}
Here, $\{n_{\mathbf{k}}\}$ represents the totality of the
vibrational quantum numbers $(....,n_{\mathbf{k}},...)$ for the
occupation of the site with position vector $\mathbf{r}_i$, and
\begin{equation}\label{energy-2}
E_b(i) = \frac{1}{N}
\sum_{\mathbf{k}}{(A_i^2/2M\omega_{\mathbf{k}}^{2})},
\end{equation}
is the small polaron binding energy. $N$ is the number of
``molecular lattice sites'' and $M$ is the appropriate reduced
atomic mass. The relation between $\omega_{\mathbf{k}}$ and its
associated wavevector \textbf{k}, i.e. the dispersion relation, is
given by:
\begin{equation}\label{dispersion}
\omega_{\mathbf{k}}^{2} =
\omega_0^2+{\omega_1^2}\sum_{\mathbf{k}}{cos({\mathbf{k}}\cdot{\mathbf{h'}})},
\end{equation}
where $|\mathbf{k}| = 2\pi p /N$, the integer $p$ lying in the
range $-(N-1)/2\leq p\leq (N-1)/2$, and $\mathbf{h}'$ indexes the
nearest neighbors $(\mathbf{r}_{i}+\mathbf{h}')$ of an arbitrary
site $\mathbf{r}_{i}$. $\omega_{0}$ is the harmonic oscillator
frequency associated with the configurational coordinate of the
isolated molecule. The relation $\omega_{1}\ll\omega_{0}$
determines the weak dispersion limit.

Equations (2) and (3) show the essential features of the GMCM which are:\\
1. site-dependent local electronic energy $\epsilon_i(0)$. \\
2. site-dependent electron-lattice interaction parameter, $A_i$,
and concomitant binding energy, $E_b(i)$.

The knowledge of $<m|V|n>$, permits the evaluation of the
``microscopic'' small polaron velocity operator
\cite{Emin:1975,Tri:1985},
\begin{equation}\label{veloc}
\mathbf{u}_{ij}=<m|\mathbf{u}|n>=(\frac{i}{\hbar})<m|V|n>(\mathbf{\mathbf{r}}_j-\mathbf{\mathbf{r}}_i),
\end{equation}
 the charge current density operator,
\begin{equation}\label{current}
\mathbf{j}_{ij}=n_{c}q\mathbf{u}_{ij},
\end{equation}
where $n_{c}$ is the charge carrier concentration, and q is the
carrier's charge, and thus the ``microscopic'' electrical
conductivity \cite{Kub:1957},
\begin{equation}\label{current}
\sigma_{ij}=\int_0^{\infty}dt\int_0^{\beta}d\rho<\mathbf{j}(-i\hbar\rho)\mathbf{j}(t)>,
\end{equation}
where $\beta=1/k_{B}T$. The mobility, $\mu_{ij}$, and consequently
the diffusion constant, given by $D_{ij}=\mu_{ij}/e\beta$, are
determined and lead to the ``microscopic'' jump rate which reads:
\begin{equation}\label{jump}
L_{ij}=\frac{D_{ij}}{|{\mathbf{r}_i-\mathbf{r}_j}|^2}
\end{equation}
Assuming that the dependence on the spatial separation $R_{ij}$,
of the two sites is \cite{Amb:1971} $\exp(-2\alpha R_{ij})$, the
``microscopic'' intrinsic transition rate, $\gamma_{ij}$, for a
small polaron hopping from a site $i$ to an empty site $j$ is
given by
\begin{equation}\label{rate}
\gamma_{ij}=exp(-2\alpha R_{ij})L_{ij}.
\end{equation}

The treatment refers to the non(anti)-adiabatic limit, i.e. in the
physical situation where the electron is no longer able to follow
rapid fluctuations of the lattice and, hence, it does not respond
quickly enough to the occurrence of a coincident event in order to
overcome the energy barrier. In this case, $J$ can be treated as a
small perturbation in the lowest order
\cite{Emin:1975,Em:1975,Alexandrov:2007}.

The expansion of the model to include the influence of possible
strong local interparticle correlations might be interesting as
intercarrier interactions exist in real systems. However, this is
beyond the aim of the present work.

\begin{figure*}[]
\includegraphics[height=6cm]{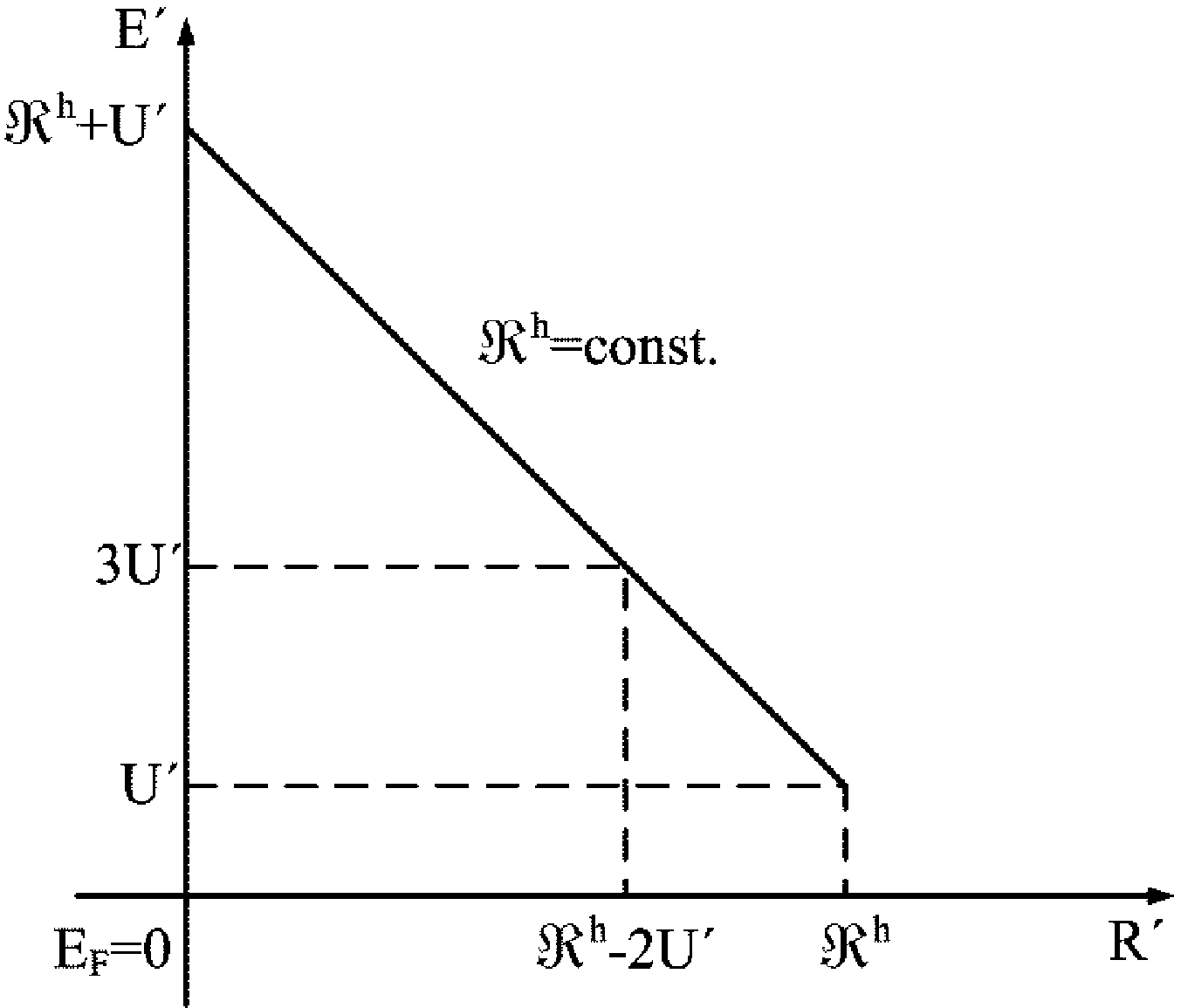}
\includegraphics[height=6cm]{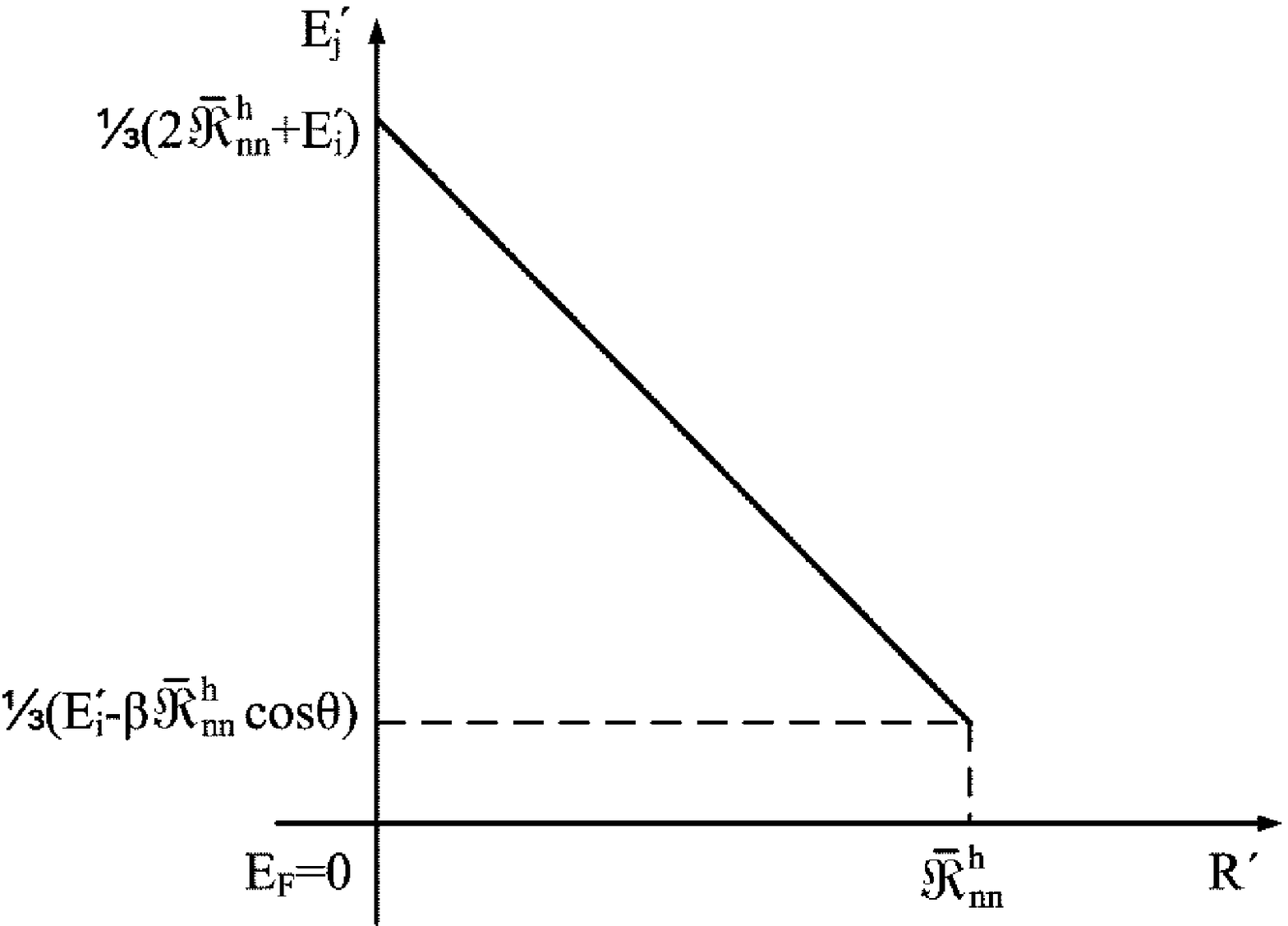}
\caption{\label{fig-Drawing1+2} \emph{High temperatures.} (I) Left panel. Contour of
constant $\Re^{h}$ for an initial site of $U'$. All possible
final sites for the carrier, $\mathcal{N}(\Re^{h})$, lie on or
within the contour $\Re^{h}$, for a particular $\theta$. (II)
Right panel. Contour of integration
$\Re^{h}=\overline{\Re}_{nn}^{h}$, for an initial site of
$E'_{i}$ and particular $\theta$, for the evaluation of
$\overline{R'}_{F}^{h}$.}
\end{figure*}

\subsection{\label{Subsec:HT} Hopping at high temperatures}
When a carrier hops from site $i$ of energy $E_{i}$ to site $j$ of
energy $E_{j}$, at a distance $R_{ij}$, the intrinsic transition
rate between the two localized states at \emph{high (h) temperatures}
($\hbar\omega_{0}\ll k_{B}T$ ~\cite{TF:1981}) is
%\begin{widetext}
\begin{equation}\label{intr.rate-high}
\gamma_{ij}^{h}=\gamma_{0}^{h}\exp(-2\alpha
R_{ij})\exp\left(-\frac{\varepsilon_{2}}{k_{B}T}\right)
\times\left\{\begin{array}{ll}
               \exp(-\frac{E_{j}-E_{i}}{2k_{B}T}), & \mbox{$E_{j}>E_{i}$} \\
               \exp( \frac{E_{i}-E_{j}}{2k_{B}T}) , & \mbox{$E_{j}<E_{i}$}
             \end{array}
      \right.
\end{equation}
%\end{widetext}
Here $\alpha^{-1}$ is the spatial extent of the electronic wave
function, $\gamma_{0}^{h} = (J^{2}/\hbar)(\pi/4\varepsilon_{2}
k_{B}T)^{1/2}$ and $\varepsilon_{2}=[E_{b}(i)+E_{b}(j)]/4$.
$E_{b}(i)$ and $E_{b}(j)$ is the small polaron binding energy for
sites $i$ and $j$, respectively. Hence, $\gamma_{ij}^{h}$ (as well
as $\gamma_{ij}^{l}$) have both spatial and energy dependence
\cite{TF:1981,TF:1986}. The spatial dimensions of the system and
the number of energies involved in the expression of the intrinsic
transition rate can be considered as the coordinates of a
``hopping space'' in which the small polaron transport occurs
under the influence of $F$. In this ``hopping space'', the most
probable hop for a carrier on a site at energy $E_{i}$ is to the
empty site at closest range, i.e. to its nearest neighbor site.
The average nearest neighbor range in the ``hopping space'',
$\overline{R}_{nn}$, determines the conductivity of the system
\cite{ApsleyHughes:1975}. Thus, to evaluate the electrical
conductivity we have to calculate this quantity first. Then,
taking into account that in real space greater real forward
distances will be hopped in the downfield direction rather than
upfield, an average real forward distance hopped should be
evaluated which, as will be presented in the following (cf.
Eqs.~\ref{mobil}-\ref{condu}), leads to the mobility of the
carriers and finally the overall conductivity of the system.

From the expression of the intrinsic transition rate between two
sites $i$ and $j$, we define the range $\Re_{ij}^{h}$ between
the sites in the ``hopping space''
\begin{equation}\label{range-high}
\Re_{ij}^{h}=2\alpha R_{ij}
+\left(\frac{\varepsilon_{2}}{k_{B}T}\right)
+\left\{\begin{array}{ll}
                \frac{E_{j}-E_{i}}{2k_{B}T}, & \mbox{$E_{j}>E_{i}$} \\
               -\frac{E_{i}-E_{j}}{2k_{B}T}, & \mbox{$E_{j}<E_{i}$}
             \end{array}
      \right.
\end{equation}
Taking the energies of the carrier to be mainly
polaronic \cite{TF:1981}, and using for convenience the terms
$E_{i}$ and $E_{j}$ instead of $E_{b}(i)$ and $E_{b}(j)$
respectively, we obtain $\varepsilon_{2}=(E_{i}+E_{j})/4$. Therefore
\begin{equation}
\Re_{ij}^{h}=2\alpha R_{ij} +\frac{3E_{j}-E_{i}}{4k_{B}T},
\end{equation}
where $E_{j}>E_{i}$ for absorption and $E_{i}/3<E_{j}<E_{i}$ for
emission of phonons. Introducing the dimensionless coordinates
$R'_{ij}=2\alpha R_{ij}$, $E'_{i}=E_{i}/2k_{B}T$ and
$E'_{j}=E_{j}/2k_{B}T$,
\begin{equation}
\Re_{ij}^{h}=R'_{ij}+\frac{3}{2}E'_{j}-\frac{1}{2}E'_{i},
\end{equation}
where $E'_{j}>E'_{i}$ for absorption and $E'_{i}/3<E'_{j}<E'_{i}$
for emission.

Under the influence of an externally applied electric field the actual energy of the hop is modified
\cite{ApsleyHughes:1975}
\begin{equation}
\frac{3}{2}E'_{j}-\frac{1}{2}E'_{i}\longrightarrow
\frac{3}{2}E'_{j}-\frac{1}{2}E'_{i}+\frac{\beta}{2}R'_{ij}\cos\theta,
\end{equation}
where $\beta = eF/2\alpha k_{B}T$ and $\theta$ is the angle
between the directions of $R'_{ij}$ and $F$. Defining the reduced
initial ($U'$) and final ($E'$) coordinates in the ``hopping
space''
\begin{equation} \label{metavlites}
U'=\frac{1}{2}E'_{i}, \; \; \; \; \;
E'=\frac{3}{2}E'_{j}+\frac{\beta}{2}R'_{ij}\cos\theta ,
\end{equation}
the range between two sites in the ``hopping space'' becomes
\begin{equation}\label{rangeHT-hopping}
\Re^{h}=R'+E'-U',
\end{equation}
where $E'>3U'$ for absorption and $U'<E'<3U'$ for emission.
The indices from $\Re_{ij}^{h}$ and $R'_{ij}$ have been dropped.

For the evaluation of the average nearest neighbor range,
$\overline{\Re}_{nn}^{h}$, firstly we have to evaluate the
number of unoccupied sites within a range $\Re^{h}$ of a
particular site of $U'$ as a function of $T$ and $F$, $\mathcal{N}(\Re^{h})$.
The three-dimensional ``hopping space'' can be represented, for a particular $\theta$ by
a two-dimensional diagram (Fig.~\ref{fig-Drawing1+2} (I)).

For hops of range less or equal to $\Re^{h}$ from an initial
site of $U'$, the final sites will lie on or within the
contour $\Re^{h}$, for a particular $\theta$, i.e. in the space
defined by $U'<E'<(\Re^{h}+U')-R'$ and $0<R'<\Re^{h}$
(Fig.~\ref{fig-Drawing1+2} (I)). Thus, using Eqs.~\ref{metavlites},
the number of empty sites enclosed by the contour $\Re^{h}$ is
\begin{equation}
\mathcal{N}(\Re^{h}) = \sum_{0,\pi} \int_{0}^{\Re^{h}}
\int_{\mathcal{LL}}^{\mathcal{UL}} N(E'_{j})[1 -
f(E'_{j})]\frac{k_{B}T}{\alpha}dE'_{j}dR'.
\end{equation}
$\mathcal{LL} = \frac{1}{3}(E'_{i}-\beta R'\cos\theta)$,
$\mathcal{UL} =
\frac{1}{3}[2\Re^{h}+E'_{i}-2R'(1+\frac{\beta}{2}\cos\theta)]$ and
$\theta=0,\pi$ as we have chosen to put $F$ exactly along the 1D
axis where transport takes place. $N(E'_{j})$ is the density of
states, and $f(E'_{j})$ the Fermi-Dirac distribution. We take the
Fermi energy $E_F=0$. Assuming a constant density of states,
$N(E^{'}_{j})=N(E^{'}_{i})=N_{0}$,
\begin{equation}
\mathcal{N}(\Re^{h}) =
\mathcal{C} \sum_{0,\pi} \int_{0}^{\Re^{h}}
\ln \frac{1+e^{\frac{2}{3}(2\Re^{h}+E'_{i}-2R'(1+\frac{\beta}{2}
\cos\theta))}}{1+e^{\frac{2}{3}(E'_{i}-\beta R'\cos\theta)}}dR',
\end{equation}
where $\mathcal{C} = \frac{N_{0}k_{B}T}{2\alpha}$.

\begin{figure*}[]
\includegraphics[height=6cm]{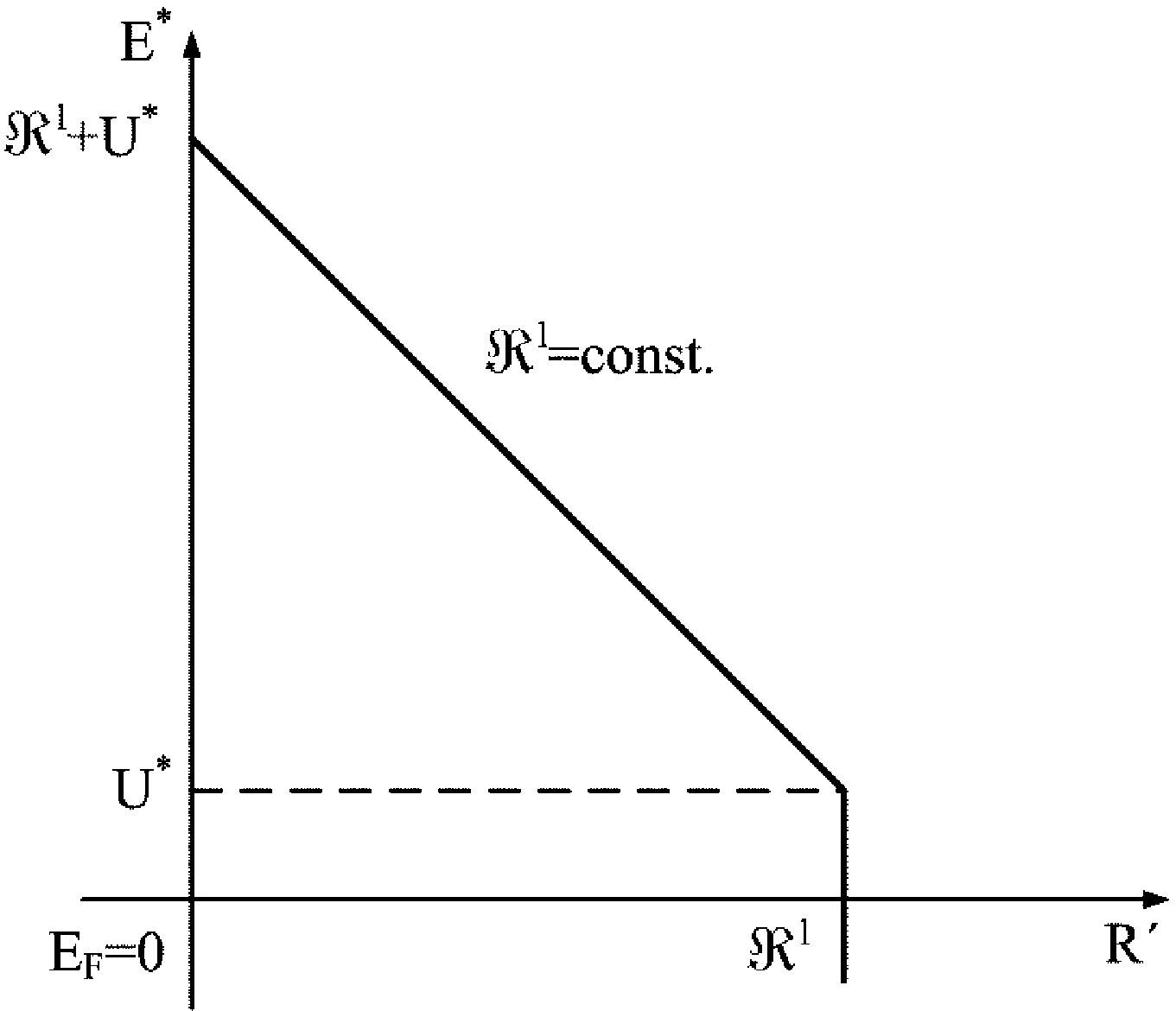}
\includegraphics[height=6cm]{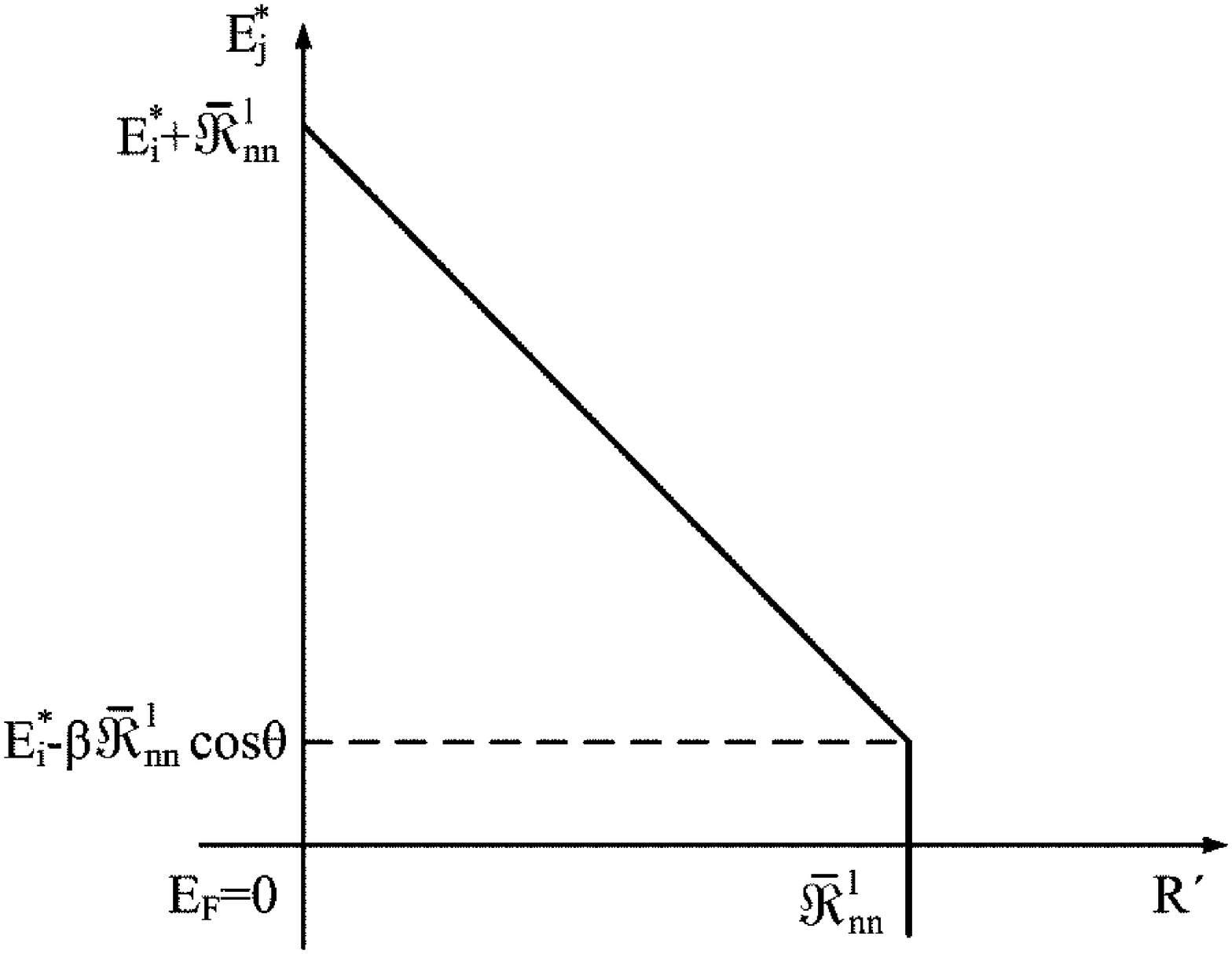}
\caption{\label{fig-Drawing3+4} \emph{Low temperatures.} (I) Left panel. Contour of
constant $\Re^{l}$ for an initial site of $U'$. All possible
final sites for the carrier, $\mathcal{N}(\Re^{l})$, lie on or
within the contour $\Re^{l}$, for a particular $\theta$. (II)
Right panel. Contour of integration
$\Re^{l}=\overline{\Re}_{nn}^{l}$, for an initial site of
$E^{*}_{i}$ and particular $\theta$, for the evaluation of
$\overline{R'}_{F}^{l}$.}
\end{figure*}

\subsection{\label{Subsec:LT} Hopping at low temperatures}
The intrinsic transition rate at \emph{low (l) temperatures}
($\hbar\omega_{0}\gg k_{B}T$ ~\cite{TF:1986}) is
%\begin{widetext}
\begin{eqnarray} \label{intr.rate-low}
\gamma_{ij}^{l}=\gamma_{0}^{l}\exp(-2\alpha R_{ij})
\left\{\begin{array}{ll}
                \exp(- \frac{E_{j}-E_{i}}{k_{B}T}), \!\! & \mbox{$E_{j}>E_{i}$}  \\
                1                                 , \!\! & \mbox{$E_{j}<E_{i}$}
             \end{array}
       \right. \! \! \! .
\end{eqnarray}
%\end{widetext}
$\gamma_{0}^{l}=
\frac{\omega_{0}}{\pi}
[ \frac{\pi J \exp(\frac{-2\varepsilon_{2}}{\hbar\omega_{0}})}{\hbar\omega_{0}}]^{2}
[(4\varepsilon_{2}/\hbar
\omega_{0})^{\Delta_{ij}/\hbar\omega_{0}}/(\Delta_{ij}/\hbar\omega_{0})!]$.
Following the same methodology as for the \emph{high temperatures}, we assign every hop of
the carrier to a hop in a three-dimensional ``hopping space''
defined by one spatial and two energy coordinates.

From the expression of the intrinsic transition rate between two
sites $i$ and $j$, we define the range
$\Re_{ij}^{l}$ between the sites in the ``hopping space''
\begin{equation} \label{range-low}
\Re_{ij}^{l}= \left\{\begin{array}{ll}
                2\alpha R_{ij} + \frac{E_{j}-E_{i}}{k_{B}T}, & \mbox{$E_{j}>E_{i}$} \\
                2\alpha R_{ij}                             , & \mbox{$E_{j}<E_{i}$}
             \end{array}
      \right. .
\end{equation}
Introducing the dimensionless coordinates $R'_{ij}=2\alpha
R_{ij}$, $E^{*}_{i}=E_{i}/k_{B}T$ and $E^{*}_{j}=E_{j}/k_{B}T$,
\begin{equation}
\Re_{ij}^{l}=\left\{\begin{array}{ll}
                      R'_{ij}+E^{*}_{j}-E^{*}_{i}, & \mbox{$E^{*}_{j}>E^{*}_{i}$} \\
                      R'_{ij}                    , & \mbox{$E^{*}_{j}<E^{*}_{i}$}
                    \end{array}
             \right. .
\end{equation}

Under the influence of an externally applied electric field the actual energy of the hop is modified \cite{ApsleyHughes:1975}
\begin{equation} E^{*}_{j}-E^{*}_{i}\longrightarrow
E^{*}_{j}-E^{*}_{i}+\beta R'_{ij}\cos\theta,
\end{equation}
where $\beta=eF/2\alpha k_{B}T$. Thus,
\begin{equation}
\Re^{l}=\left\{\begin{array}{ll}
                      \!\!\! R'(1+\beta\cos\theta)+E^{*}_{j}-E^{*}_{i}, & \! \mbox{$E^{*}_{j}>E^{*}_{i}-\beta R'\cos\theta$} \\
                      \!\!\! R'                                       , & \! \mbox{$E^{*}_{j}<E^{*}_{i}-\beta R'\cos\theta$}
                    \end{array}
             \right. \!\! .
\end{equation}
The indices from $\Re_{ij}^{l}$ and $R'_{ij}$ have been dropped.
Defining the reduced initial ($U'$) and final ($E'$) coordinates in the ``hopping space''
\begin{equation} \label{metavliteslow}
U^{*}=E^{*}_{i}, \; \; \; \; \; E^{*}=E^{*}_{j}+\beta
R'\cos\theta,
\end{equation}
the range between two sites in the ``hopping space'' becomes
\begin{equation} \label{rangeLT-hopping}
\Re^{l}=\left\{\begin{array}{ll}
                      R'+E^{*}-U^{*}, & \mbox{$E^{*}>U^{*}$} \\
                      R'            , & \mbox{$E^{*}<U^{*}$}
                    \end{array}
             \right. \!\! .
\end{equation}
For the evaluation of the average nearest neighbor range in the
``hopping space'', $\overline{\Re}_{nn}^{l}$, firstly we have to
evaluate the number of unoccupied sites within a range $\Re^{l}$
of a particular site of $U^{*}$ as a function of
$T$ and $F$, $\mathcal{N}(\Re^{l})$. The
``hopping space'' can be represented, for a particular $\theta$ by
a two-dimensional diagram (Fig.~\ref{fig-Drawing3+4}(I)).

For hops of range less or equal to $\Re^{l}$ from an initial
site of $U^{*}$, the final sites will lie on or within the
contour $\Re^{l}$, for a particular $\theta$, i.e. in the space
defined by $-\infty<E^{*}<(\Re^{l}+U^{*})-R'$ and
$0<R'<\Re^{l}$ (Fig.~\ref{fig-Drawing3+4} (I)). Thus, using
Eqs.~\ref{metavliteslow}, the number of empty sites
enclosed by the contour $\Re^{l}$ is
\begin{equation}
\mathcal{N}(\Re^{l}) = \sum_{0,\pi} \int_{0}^{\Re^{l}}
\int_{\mathcal{LL}}^{\mathcal{UL}} N(E^{*}_{j})[1 -
f(E^{*}_{j})]\frac{k_{B}T}{2\alpha}dE^{*}_{j}dR'.
\end{equation}
$\mathcal{LL} = -\infty$,
$\mathcal{UL} = \Re^{l}+E^{*}_{i}-R'(1+\beta\cos\theta)$.
Assuming a constant density of states,
$N(E^{*}_{j})=N(E^{*}_{i})=N_{0}$,
\begin{equation}
\mathcal{N}(\Re^{l}) = \mathcal{C} \sum_{0,\pi} \int_{0}^{\Re^{l}}
\ln[1+\exp(\Re^{l}+E^{*}_{i}-R'(1+\beta \cos\theta))]dR',
\end{equation}
where again $\mathcal{C} = \frac{N_{0}k_{B}T}{2\alpha}$.

\subsection{\label{Subsec:conductivity} Conductivity}
We define $E''_{i} = E'_{i}$ for \emph{high temperatures} or
$E''_{i} = E^{*}_{i}$ for \emph{low temperatures}.
The knowledge of the number of unoccupied sites within a range $\Re^{h/l}$
for either \emph{high} or \emph{low} ($h/l$) \emph{temperatures},
$\mathcal{N}(\Re^{h/l})$,
permits the evaluation of the average nearest neighbor range, $\overline{\Re}_{nn}^{h/l}$,
when the carrier resides on a particular site of $E''_{i}$,
as a function of $T$ and $F$ \cite{ApsleyHughes:1975}
\begin{equation}
\overline{\Re}_{nn}^{h/l} = \int_{0}^{\infty}\Re^{h/l}
\frac{\partial\mathcal{N}(\Re^{h/l})}{\partial\Re^{h/l}}\exp[-\mathcal{N}(\Re^{h/l})]d\Re^{h/l},
\end{equation}
or equivalently
\begin{equation}
\overline{\Re}_{nn}^{h/l}=\int_{0}^{\infty}\exp[-\mathcal{N}(\Re^{h/l})]d\Re^{h/l}.
\end{equation}
The evaluation of $\overline{\Re}_{nn}^{h/l}$,
gives the range in the three-dimensional ``hopping space''
where a nearest neighbor exists that can host the carrier
when the carrier hops from an initial site of $E''_{i}$.
However, it gives no information on the direction of the hop of the carrier.

Considering, all sites of initial $E''_{i}$ and
assuming that all hops from these sites are all hops of range
$\overline{\Re}_{nn}^{h/l}$, then in real space, these hops will
be in random directions, but for a hop to final sites of the same
energy, greater real forward distance will be hopped in the
downfield direction rather than upfield. Thus, summing over all
final sites, for initial sites of $E''_{i}$, there will be
associated an average real forward distance hopped \cite{ApsleyHughes:1975}
\begin{equation}
\overline{R}_{F}^{h/l}=\frac{\overline{R'}_{F}^{h/l}}{2\alpha}.
\end{equation}

For \emph{high temperatures},
the distance $\overline{R}_{F}^{h}$ is evaluated
by averaging
$R'\cos\theta$ over the contour
$\overline{\Re}_{nn}^{h} = const.$ (Fig.~\ref{fig-Drawing1+2} (II)), and hence
\begin{equation}
\overline{R'}_{F}^{h}=\frac{I_{1}}{I_{2}}.
\end{equation}
The integrals $I_{1}$ and $I_{2}$ are given in the Appendix.

For \emph{low temperatures},
the distance $\overline{R}_{F}^{l}$ is evaluated by averaging
$R'\cos\theta$ over the contour $\overline{\Re}_{nn}^{l} =
const.$ (Fig.~\ref{fig-Drawing3+4} (II)), and hence
\begin{equation}
\overline{R'}_{F}^{l}=\frac{I_{1}+I_{2}}{I_{3}+I_{4}}.
\end{equation}
The integrals $I_{1}$, $I_{2}$, $I_{3}$, $I_{4}$ are given in the Appendix.

Either for \emph{high} or for \emph{low} \emph{temperatures},
having calculated the distance $\overline{R'}_{F}^{h/l}$ and
considering that the probability of all hops is
$\exp(-\overline{\Re}_{nn}^{h/l})$, the average rate of
transport of carriers is
$\nu_{ph}\overline{R'}_{F}^{h/l}\exp(-\overline{\Re}_{nn}^{h/l})$.
Here, $\nu_{ph}$ is a hopping attack frequency of the order of a
phonon frequency, assumed the same for all hops.

The mobility for small polarons of $E''_{i} = E'_{i}$ for
\emph{high temperatures} or of $E''_{i} = E^{*}_{i}$ for \emph{low temperatures}
reads
\begin{equation}\label{mobil}
\mu(E''_{i})=-\frac{\nu_{ph}}{F}\frac{\overline{R'}_{F}^{h/l}}{2\alpha}\exp(-\overline{\Re}_{nn}^{h/l}),
\end{equation}
and the conductivity of the system is
\begin{equation} \label{condu}
\sigma^{h/l}(F,T) = - \!\! \int_{-\infty}^{\infty} \! \! \! \! \!
\! eN(E''_{i})f(E''_{i})\mu(E''_{i}) k_{B}TdE''_{i}.
\end{equation}

\section{Results and discussion}\label{Sec:resdis}
In the following, based on the theoretical analysis presented
above, we calculate numerically the electrical conductivity
varying the density of states and the spatial extent of the
localized electronic wave function. Our numerical results refer
for simplicity reasons to a constant density of states, although,
typically, in 1D systems the density of states has a strong energy
dependence. One could alternatively use an energy dependent model
for the density of states \cite{TZYK:1991,Triberis:1992} which is
expected to influence somehow the conductivity. This is beyond the
scope of the present paper, but could be numerically examined in
the future via the same approach, as it is evident from
Eq.~\ref{condu}.

We consider the range $T=(160$ - $300)$ K as \emph{high temperatures} and
the range $T=(10$ - $150)$ K as \emph{low temperatures}.
We investigate the influence of
an electric field in the range
$F=(5\times 10^{3}$ - $1\times 10^{8})$ Vm$^{-1}$.
We did not consider higher values of $F$,
because it is generally accepted that the value of
the highest electric field that the polaron can sustain is about
$1 \times 10^8$ Vm$^{-1}$
\cite{RakhmanovaConwell:1999,Liu:2006,QiuZhu:2009}.
We take $\nu_{ph}=10^{12}$ s$^{-1}$.

We vary the density of states by orders of magnitude around values
which are relevant to common 1D
systems~\cite{HKS:2010,HKS:2011:erratum,Bourbie:2007,Bourbie:2011}
and the extent of the electronic wave function from 1 to 5 $\AA$,
i.e. for sensible values for common organic
molecules~\cite{HKS:2009,HSK:2009}. Specifically, in
Figs.~\ref{fig-HT-ALLDOS},~\ref{fig-HT-compF},~\ref{fig-HT-beta9+11}
and
Figs.~\ref{fig-LT-T12},~\ref{fig-LT-Fa-Fc},~\ref{fig-LT-beta9+11}
we keep $\alpha^{-1} =$ 2 $\AA$, while in
Figs.~\ref{fig-HT-alphaTF},~\ref{fig-HT-alphabeta-linear} we vary
$\alpha^{-1}$ in the range 1 - 5 $\AA$. However, to keep our
results as general as possible, we do not make any reference to a
specific material.

\subsection{High temperatures} \label{ResSubsec:HT}
\begin{figure*}[h!]
\includegraphics[width=9cm]{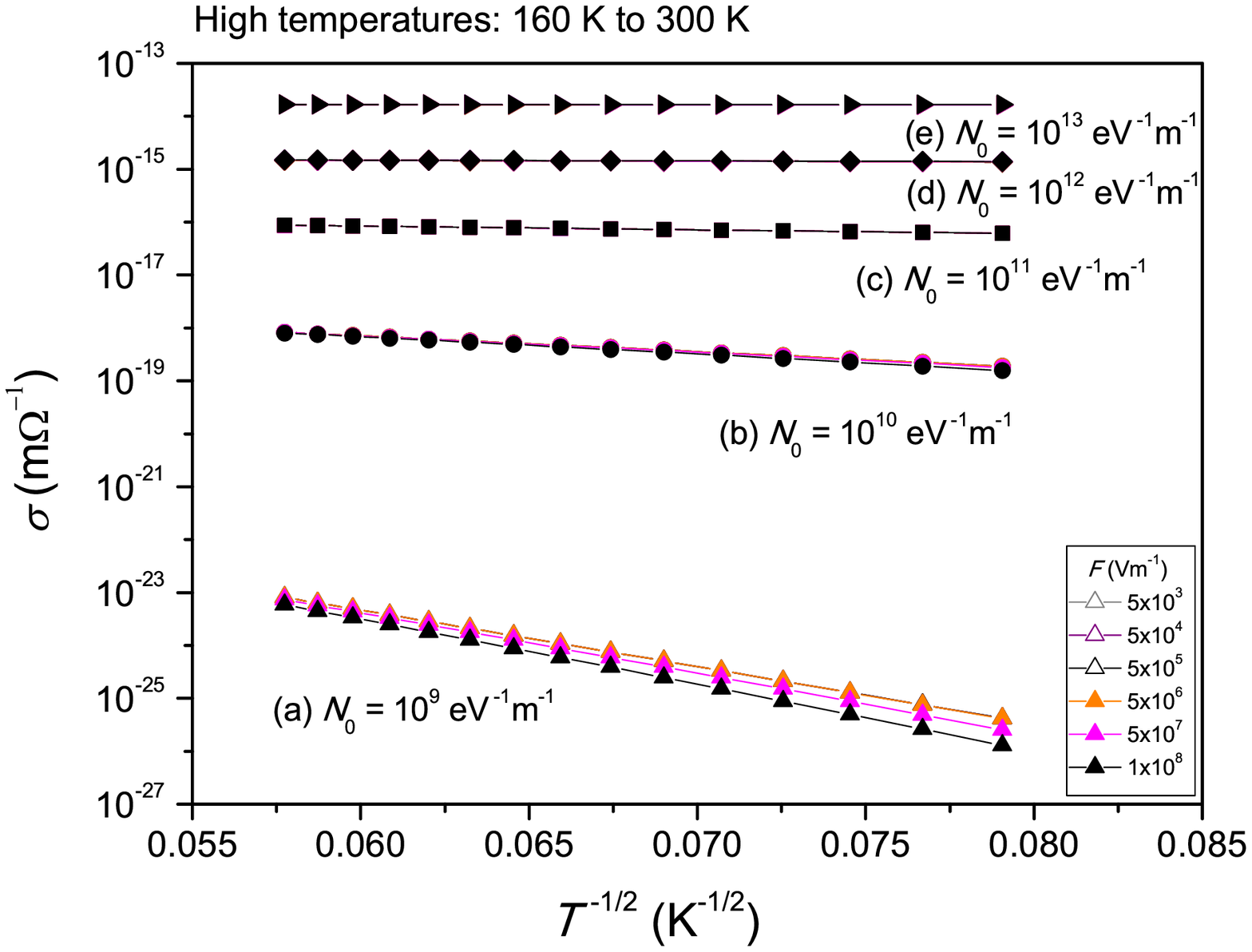}
\hspace*{-1cm}
\includegraphics[width=9cm]{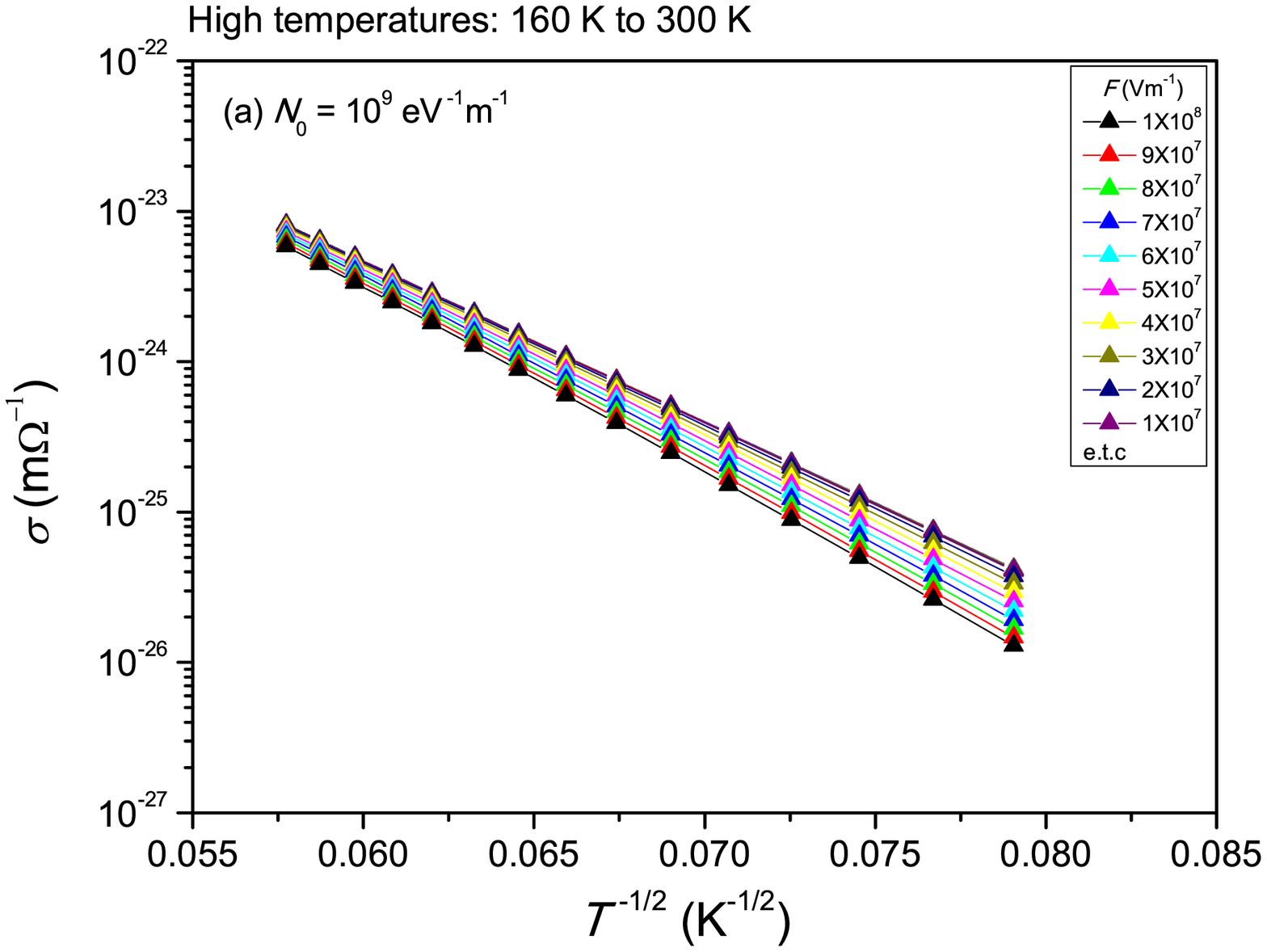}
\includegraphics[width=9cm]{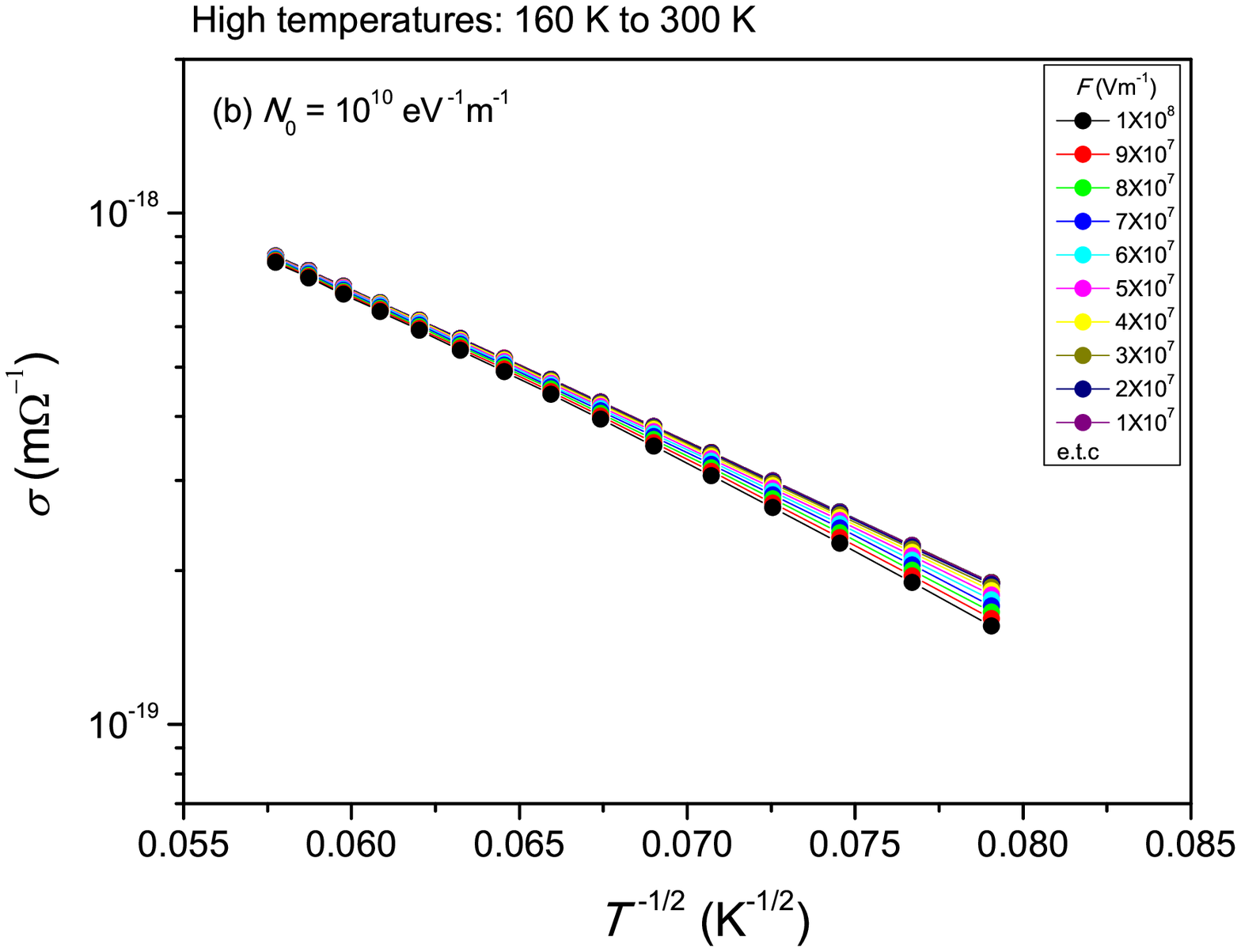}
\hspace*{-1cm}
\includegraphics[width=9cm]{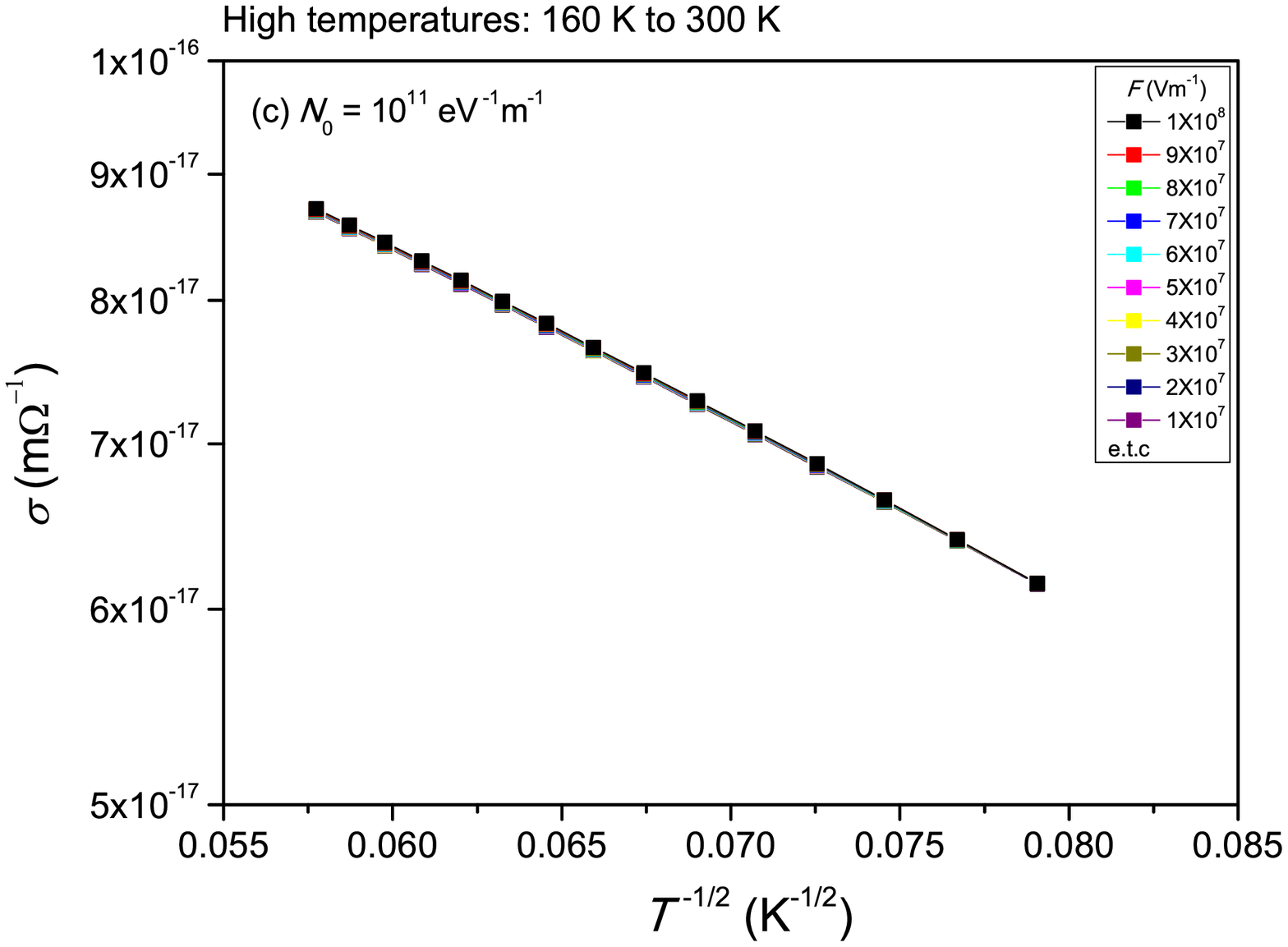}
\includegraphics[width=9cm]{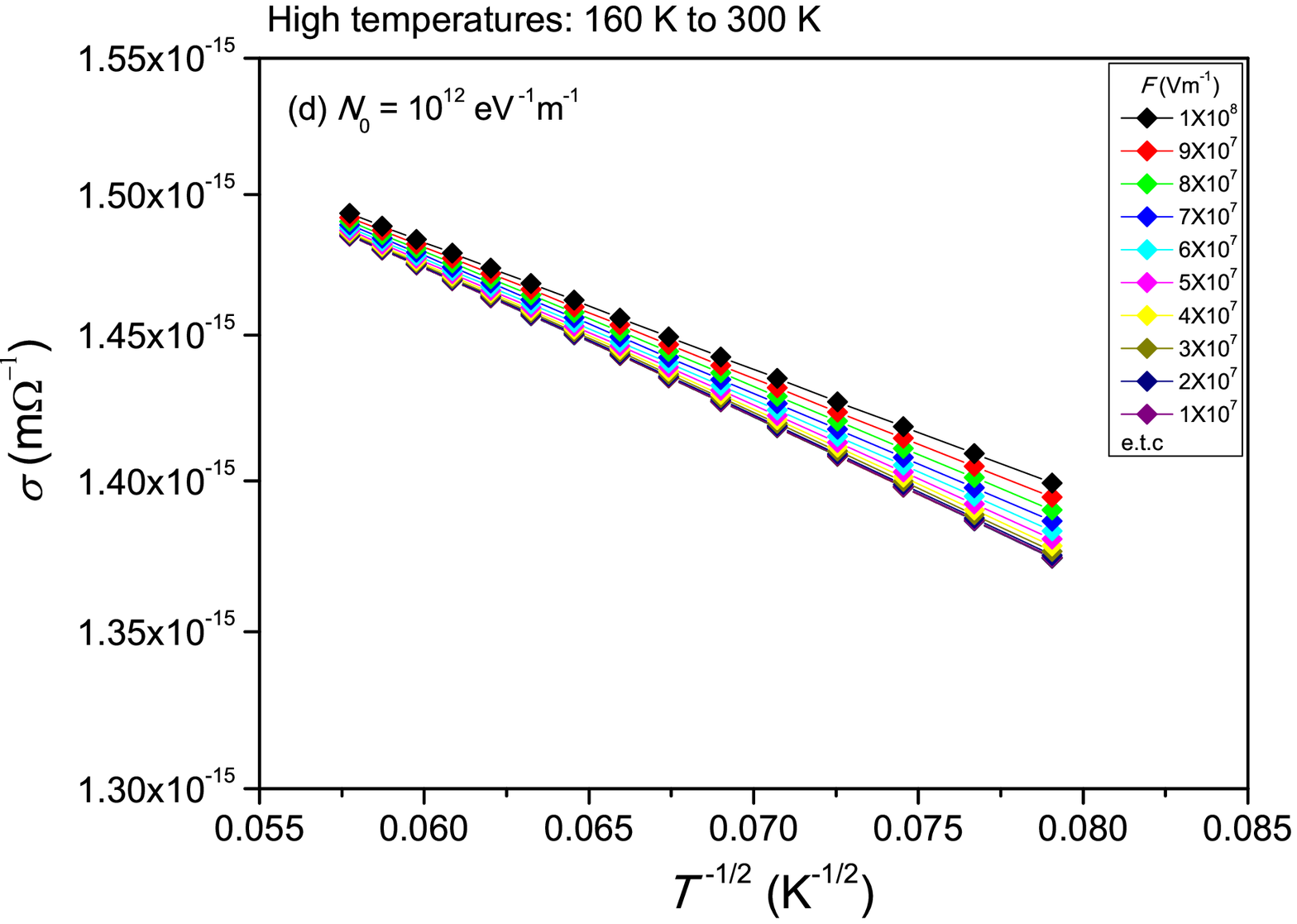}
\hspace*{-1cm}
\includegraphics[width=9cm]{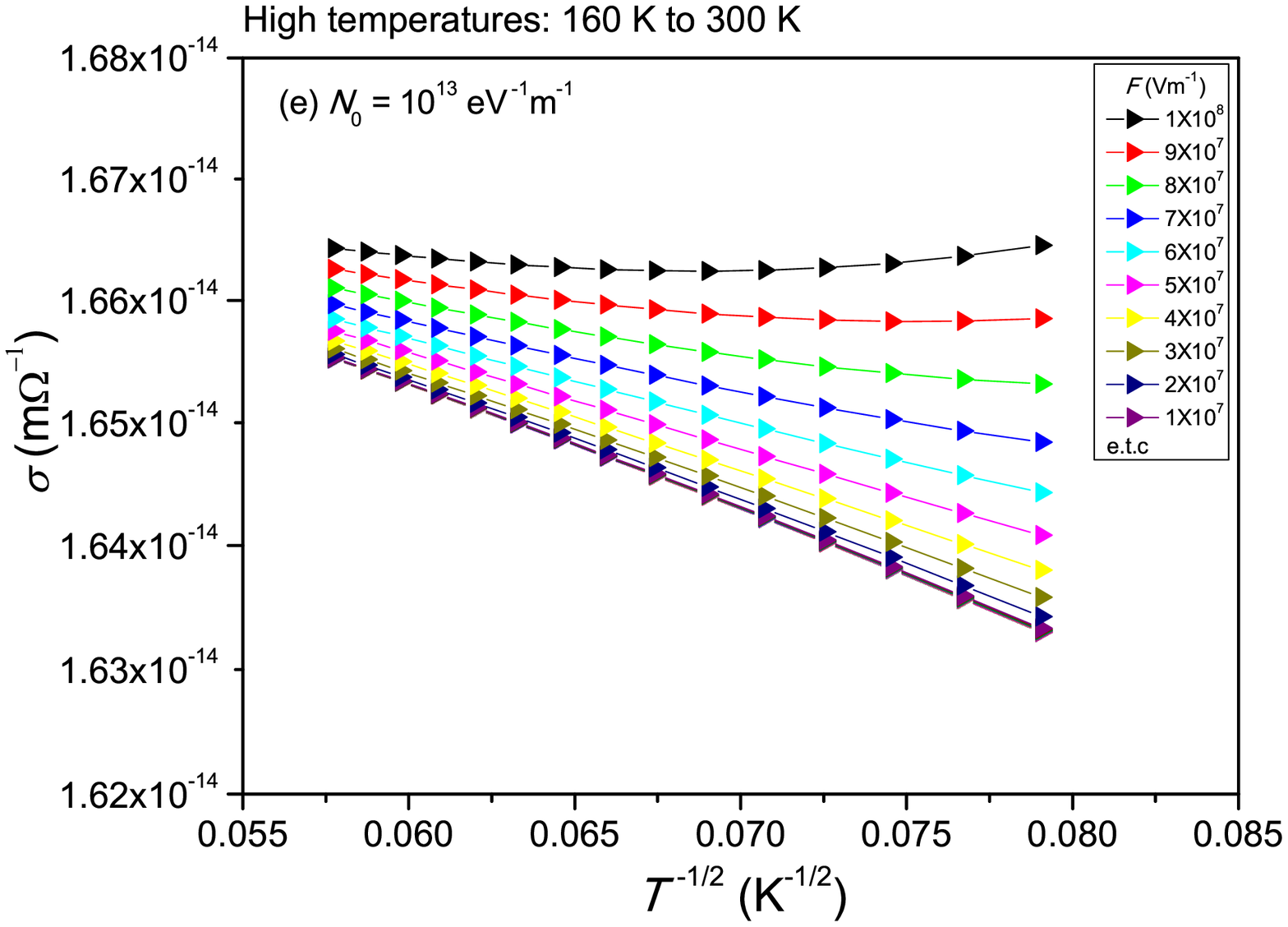}
\vspace*{-0.5cm} \caption{\label{fig-HT-ALLDOS}
$\sigma$ versus $T^{-1/2}$, varying the density of states i.e. for
$N_{0}=10^{9}$ - $10^{13}$ eV$^{-1}$m$^{-1}$ [cases (a) to (e), respectively].
$\alpha^{-1}=2$ $\AA$.
$T=(160$ - $300)$ K
and $F=(5\times 10^{3}$ - $1\times 10^{8})$ Vm$^{-1}$.}
\end{figure*}
For the temperature range $T=(160$ - $300)$ K,
Fig.~\ref{fig-HT-ALLDOS} presents
$\sigma$ as a function of $T^{-1/2}$ for
different magnitudes of the DOS, i.e. for
$N_{0}=(10^{9}$ - $10^{13})$ eV$^{-1}$m$^{-1}$ [cases (a) to (e), respectively].
$\alpha^{-1}=2$ $\AA$.
$F=(5 \times 10^{3}$ - $1\times 10^{8})$ Vm$^{-1}$.
We depict our results as a function of $T^{-1/2}$
to compare them with the analytically obtained formula
$\ln\sigma\propto T^{-1/2}$ which holds for low up to moderate
electric fields and was previously obtained by two of us
\cite{DT:2010:Fcr} following a different theoretical treatment and
taking into account the effect of correlations, namely
\begin{equation}
\sigma^{h,cr}(F,T) \propto
\exp\left[-\left(\frac{T_{0}^{h,cr}}{T}\right)^{1/2} \!\!\!\!\!\! \left(1-\frac{F^{2}}{g(T)}\right)^{1/2}\right].
\end{equation}
Here $g(T)=\left(2\alpha k_{B}T/e\right)^{2}$ and
$T_{0}^{h,cr}=1.18\alpha/k_{B}N_{0}$.  We observe that higher density of
states leads to higher conductivity in such a way that
$\frac{d\sigma}{dN_0}$ is smaller for higher densities of states.
As shown in Fig.~\ref{fig-HT-ALLDOS}, in this specific example,
augmenting DOS by four orders of magnitude,
the conductivity rises by approximately eleven orders of magnitude.

\begin{figure}[]
\includegraphics[width=9cm]{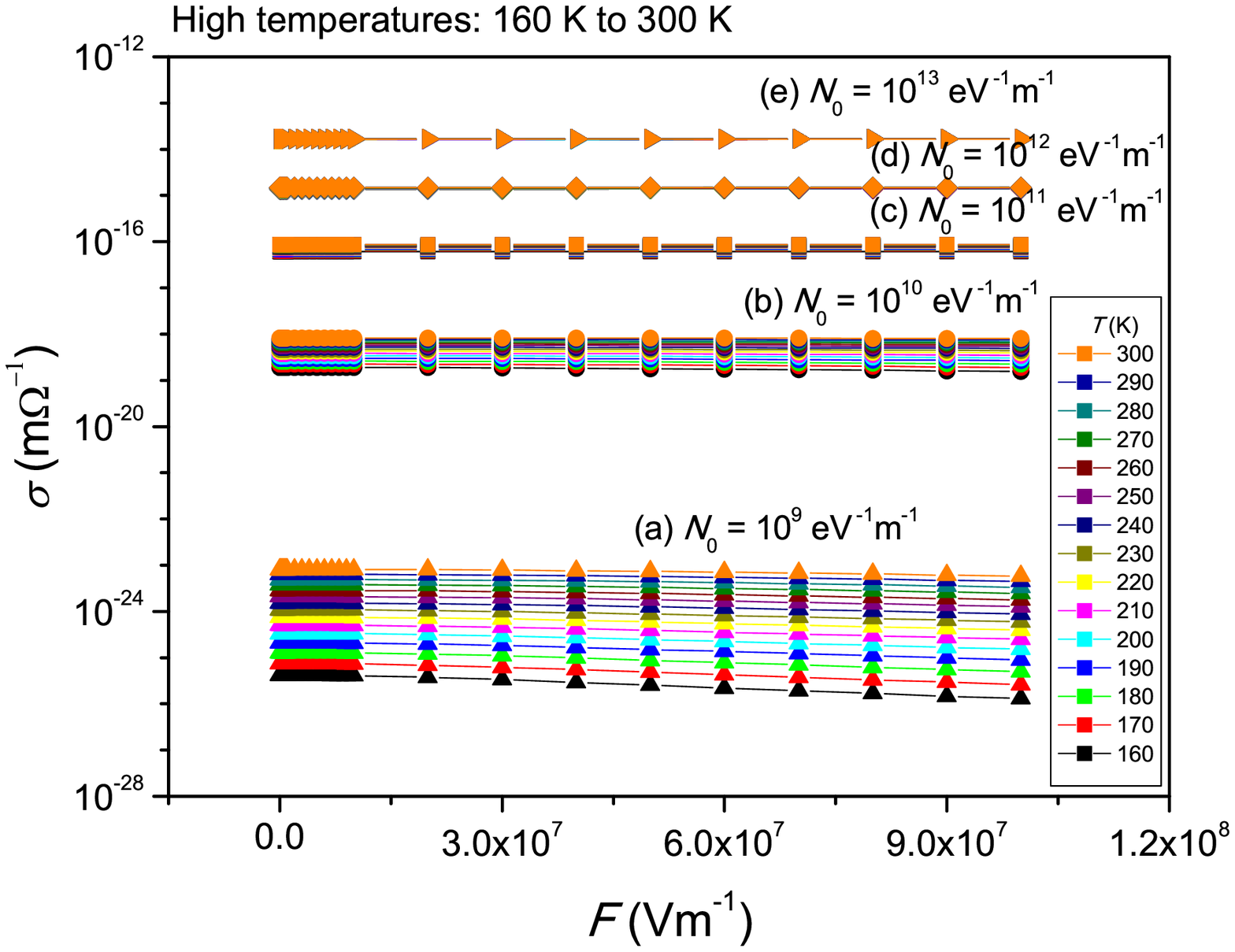}
\hspace*{-1cm}
%\vspace*{-0.5cm}
\caption{\label{fig-HT-compF}
$\sigma$ versus $F$, varying the density of states i.e.
for $N_{0}=10^{9}$ - $10^{13}$ eV$^{-1}$m$^{-1}$
[cases (a) to (e), respectively].
$\alpha^{-1}=2$ $\AA$.
$T=(160$ - $300)$ K and $F=(5\times 10^{3}$ - $1\times10^{8})$ Vm$^{-1}$.}
\end{figure}

From these results we realize that the $T^{-1/2}$-behaviour of
$\sigma$ \cite{DT:2010:Fcr} holds for low up to moderate electric fields.
For the smaller densities of states, i.e. for
$N_{0}=10^{9}$ and $10^{10}$ eV$^{-1}$m$^{-1}$,
$\frac{d\sigma}{dF} < 0$ i.e. the conductivity is larger for lower
electric field strengths. This is due to the competitive role of
the directionality imposed by the electric field and the
temperature. This directionality affects destructively $\sigma$
when not many sites are available for the carrier i.e. for small
densities of states. Here we notice that according to Eq.~\ref{rangeHT-hopping} and
Fig.~\ref{fig-Drawing1+2}(I),
the electric field affects the range between two sites in the ``hopping
space'' both for the absorption and the emission branch. On the
contrary, this effect does not appear at \emph{low temperatures}
($T=(10$ - $150)$ K), discussed later on
Subsection~\ref{ResSubsec:LT} (Fig.~\ref{fig-LT-T12}) because in
the corresponding expression for the range between two sites in
the ``hopping space'' at \emph{low temperatures} (Eq.~\ref{rangeLT-hopping} and Fig.~\ref{fig-Drawing3+4}(I)),
the electric field affects only the finite area of the absorption
branch. We mention that the electric field plays a constructive
role, too, due to its energy offer to the carriers. At
$N_{0}=10^{11}$ eV$^{-1}$m$^{-1}$ it seems that the available
sites are numerous enough so that the directionality of the
electric field hardly affects the conductivity. For higher
densities of states, i.e. for $N_{0}=10^{12}$ and $10^{13}$
eV$^{-1}$m$^{-1}$, only the constructive energetic influence of
the electric field appears. Now $\frac{d\sigma}{dF} > 0$ i.e. the
conductivity is larger for higher electric field strengths.
Finally, we observe that the temperature has a greater effect on
the conductivity, the smaller the density of states is. Another
aspect of the behaviour of the conductivity for different DOS is
shown in Figure \ref{fig-HT-compF}. $N_{0}=10^{9}$ - $10^{13}$
eV$^{-1}$m$^{-1}$, $T=(160$ - $300)$ K and $F=(5\times 10^{3}$ -
$1\times 10^{8})$ Vm$^{-1}$.

\begin{figure*}[]
\includegraphics[width=9cm]{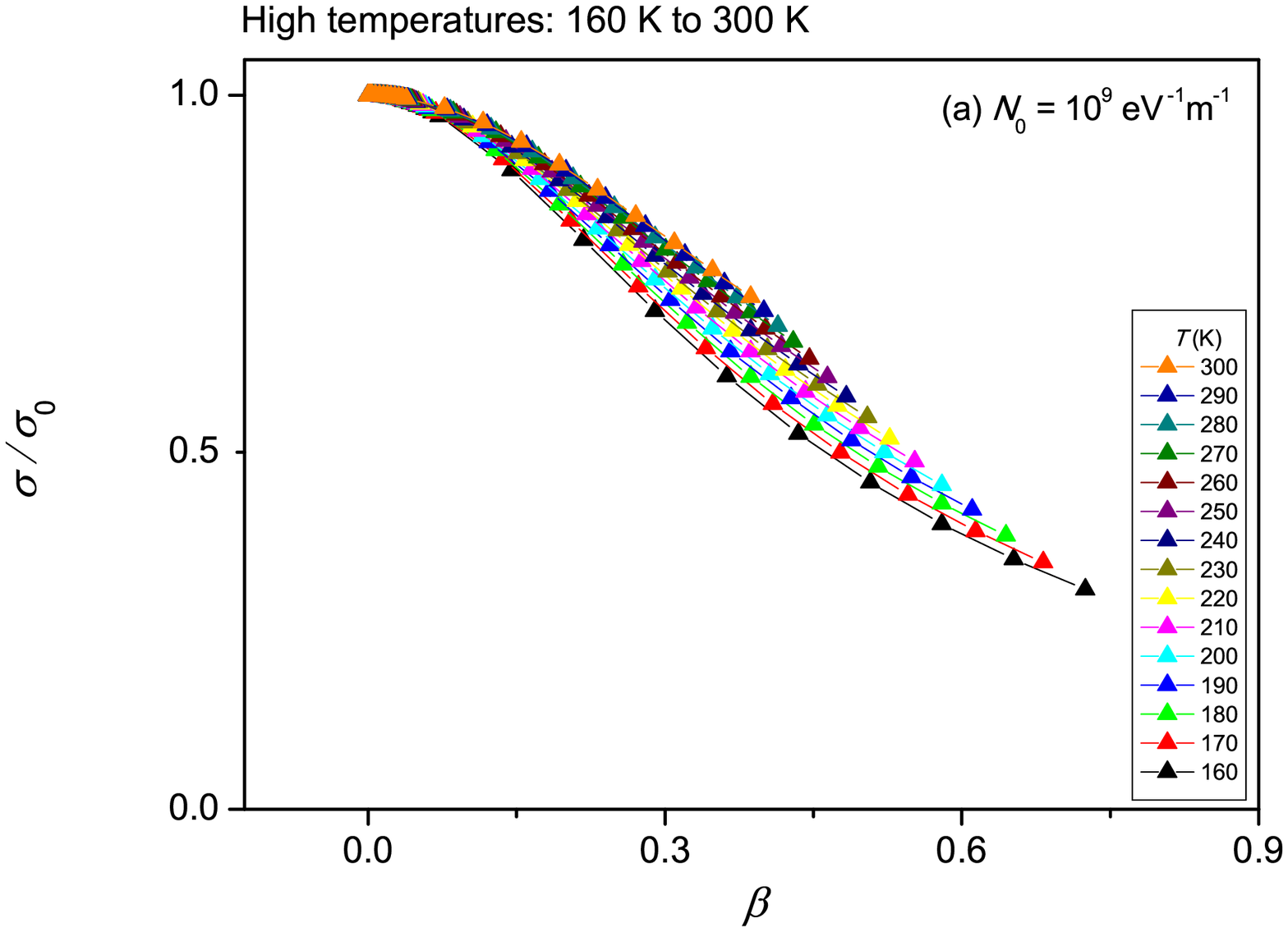}
\hspace*{-1cm}
\includegraphics[width=9cm]{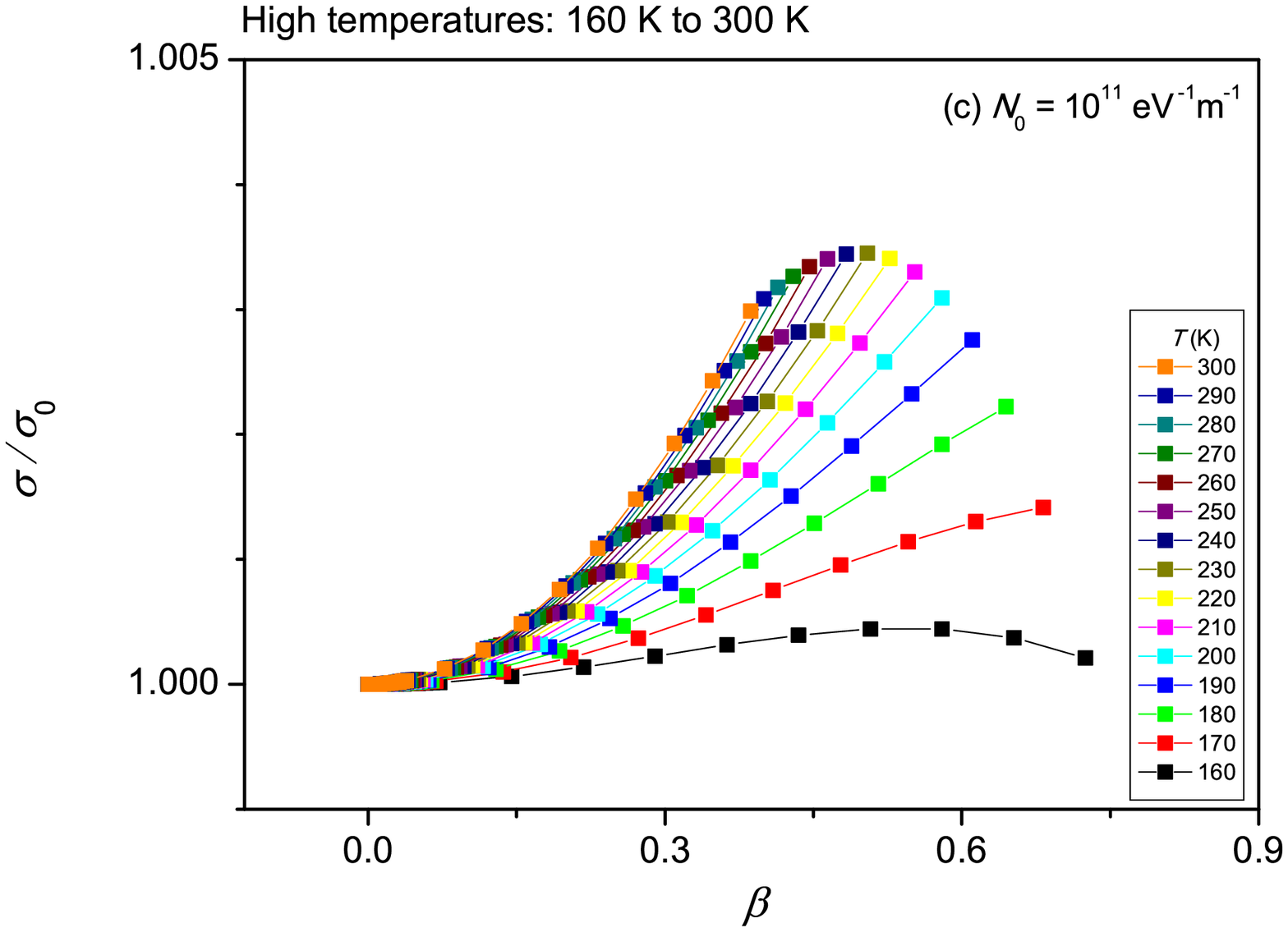}
\vspace*{-0.5cm}
\caption{\label{fig-HT-beta9+11}
$\sigma/\sigma_{0}$ versus $\beta = eF/2\alpha k_{B}T$.
(I) Left Panel. Case (a) $N_{0}=10^{9}$ eV$^{-1}$m$^{-1}$.
(II) Right Panel. Case (c) $N_{0}=10^{11}$ eV$^{-1}$m$^{-1}$.
In both panels $\alpha^{-1}=2$ $\AA$,
$F=(5\times 10^{3}$ - $1\times 10^{8})$ Vm$^{-1}$ and
$T=(160$ - $300)$ K.}
\end{figure*}

Let us denote by $\sigma_{0}$ the ohmic value of the conductivity,
i.e. $\sigma_{0} = \lim_{F\to 0}(\sigma) $. In order to show the
deviation of $\sigma$ from $\sigma_{0}$ under the influence of
both $F$ and $T$ we present Fig.~\ref{fig-HT-beta9+11} which shows
$\sigma/\sigma_{0}$ versus $\beta = eF/2\alpha k_{B}T$ for (a)
$N_{0}=10^{9}$ eV$^{-1}$m$^{-1}$ and (c) $N_{0}=10^{11}$
eV$^{-1}$m$^{-1}$, respectively. $F=(5\times 10^{3}$ - $1\times
10^{8})$ Vm$^{-1}$ and $T=(160$ - $300)$ K. We observe that the
effect of the external stimuli $F$ and $T$ on $\sigma$ depends
strongly on the value of the density of states that characterizes
the system. For (a) $N_{0}=10^{9}$ eV$^{-1}$m$^{-1}$ the
conductivity decreases from its ohmic value in the specific range
of $F$ and $T$, while for (c) $N_{0}=10^{11}$ eV$^{-1}$m$^{-1}$
the conductivity generally increases and it is higher for higher
temperatures. We notice that the variation of $\sigma/\sigma_{0}$
versus $\beta = eF/2\alpha k_{B}T$ is generally small especially
in contrast to the corresponding variation at \emph{low temperatures}
studied later on Subsection~\ref{ResSubsec:LT}.

\begin{figure*}[]
\includegraphics[width=9cm]{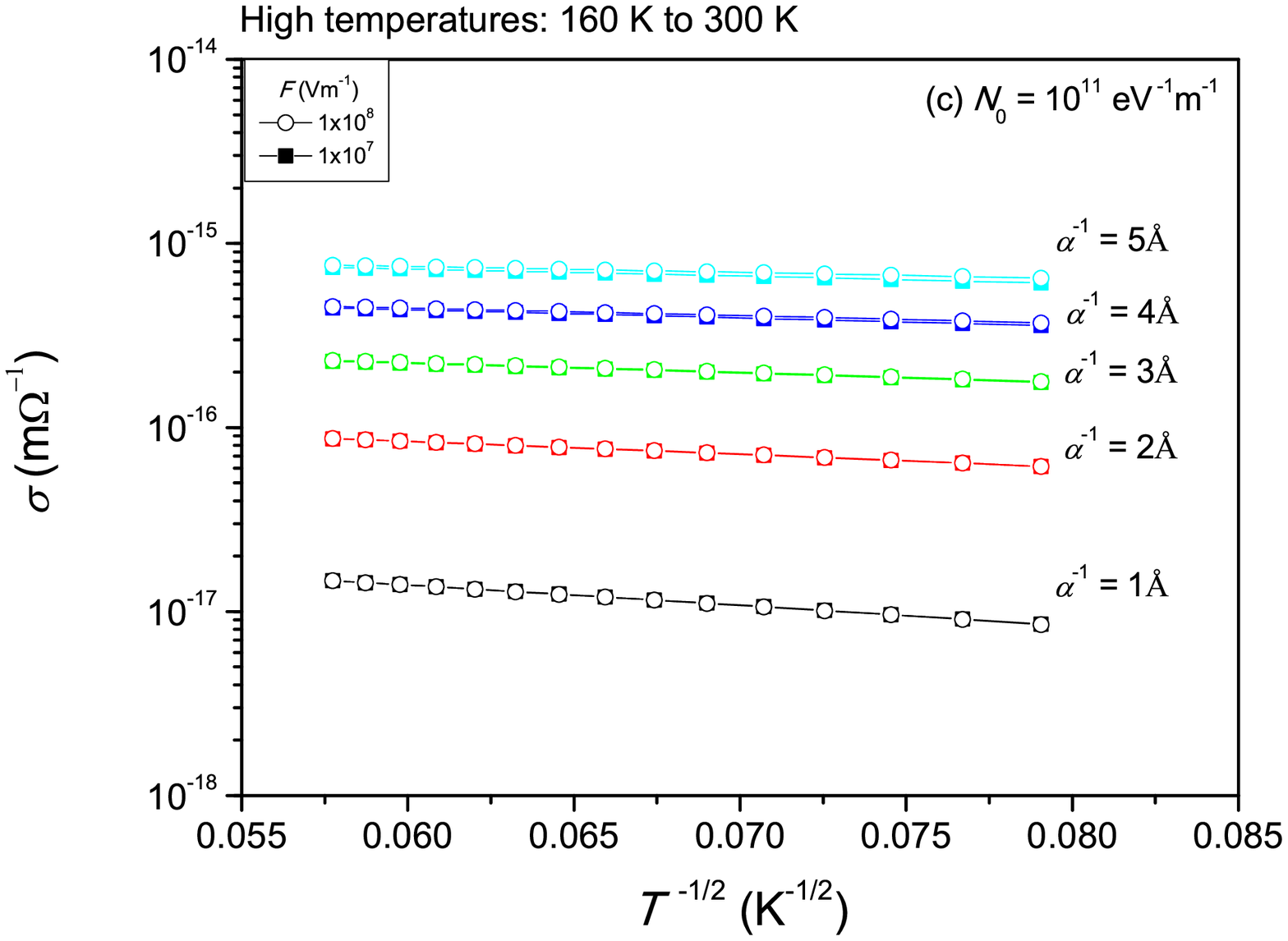}
\hspace*{-1cm}
\includegraphics[width=9cm]{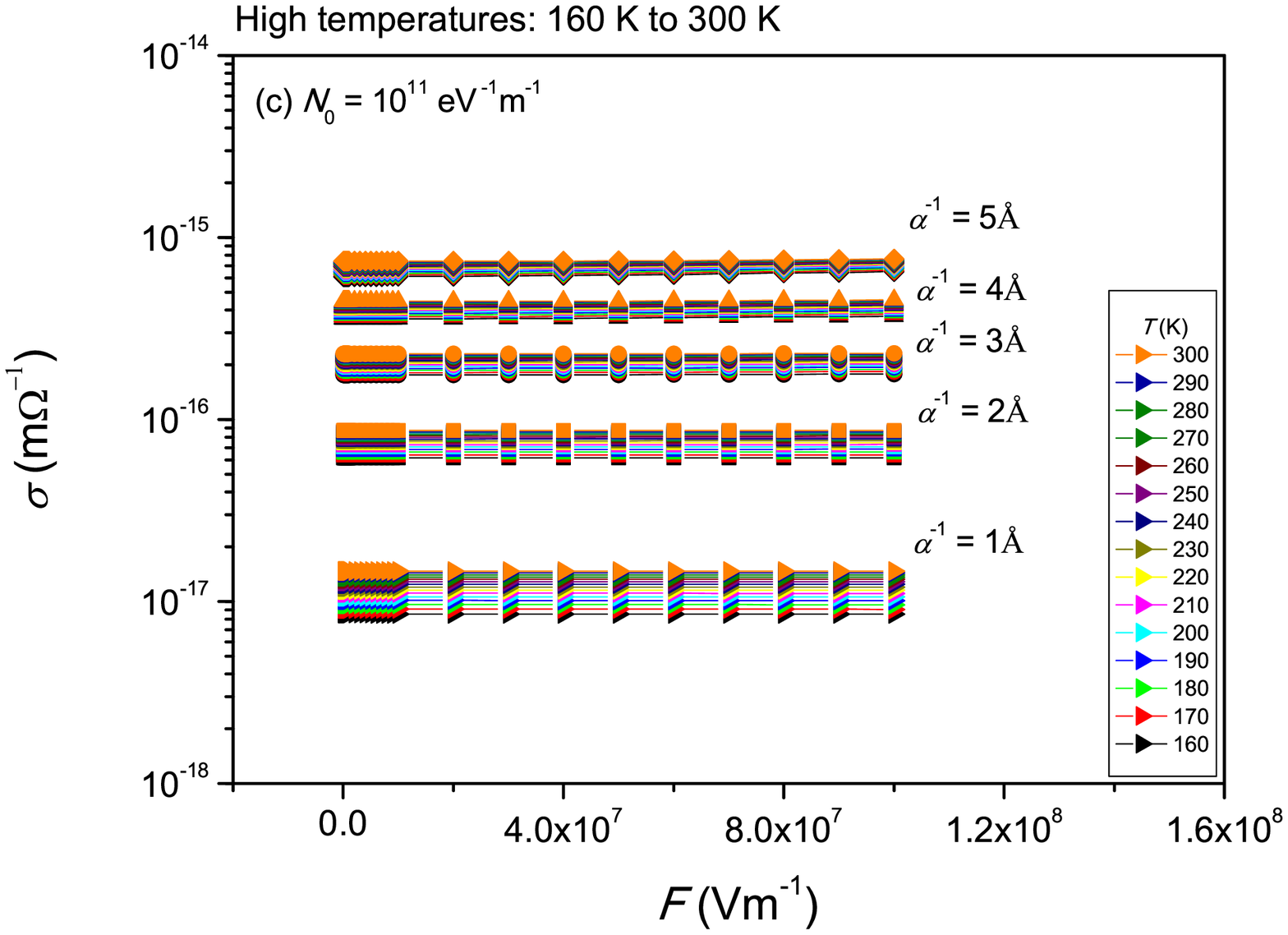}
\hspace*{+4cm} \vspace*{-0.5cm} \caption{\label{fig-HT-alphaTF}
(I) Left Panel. $\sigma$ versus $T^{-1/2}$. (II) Right Panel.
$\sigma$ versus $F$. The spatial extent of the localized
electronic wave function is taken $\alpha^{-1}=(1$ - $5)$ $\AA$.
Here we have chosen case (c) $N_{0}=10^{11}$ eV$^{-1}$m$^{-1}$ for
the density of states. $T=(160$ - $300)$ K and $F=(5\times 10^{3}$
- $1\times 10^{8})$ Vm$^{-1}$.}
\end{figure*}

\begin{figure}[]
\includegraphics[width=9cm]{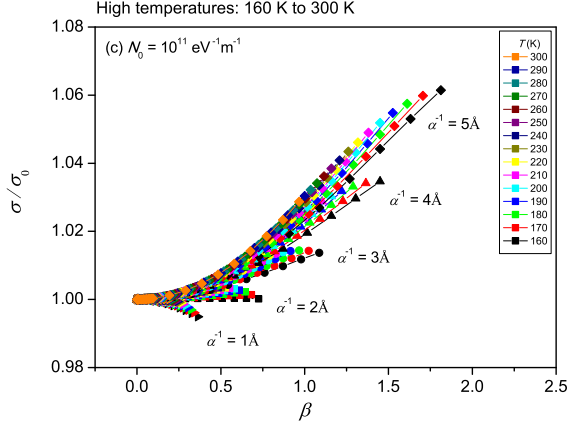}
\hspace*{-1cm} %\vspace*{-0.7cm}
\caption{\label{fig-HT-alphabeta-linear}
$\sigma/\sigma_{0}$ versus $\beta = eF/2\alpha k_{B}T$ for
$\alpha^{-1}=(1$ - $5)$ $\AA$. For the density of states we have
chosen (c) $N_{0}=10^{11}$ eV$^{-1}$m$^{-1}$. $T=(160$ - $300)$ K
and $F=(5\times 10^{3}$ - $1\times 10^{8})$ Vm$^{-1}$.}
\end{figure}

Figure \ref{fig-HT-alphaTF} shows the conductivity for different
values of the spatial extent of the localized electronic wave
function, i.e. for $\alpha^{-1}=(1$ - $5)$ $\AA$. Here we have
chosen case (c) $N_{0}=10^{11}$ eV$^{-1}$m$^{-1}$ for the density
of states. $T=(160$ - $300)$ K and $F=(5\times 10^{3}$ - $1\times
10^{8})$ Vm$^{-1}$. We observe that smaller $\alpha^{-1}$ (more
localized carriers) leads to smaller $\sigma$. Particularly, five
times increase of $\alpha^{-1}$ leads to two orders of magnitude
greater conductivity. In addition, in
Fig.~\ref{fig-HT-alphabeta-linear} we observe that
$\frac{d\sigma}{d\beta}>0$ for any temperature when $\alpha^{-1}
=$ 3, 4 and 5 $\AA$, while $\frac{d\sigma}{d\beta} < 0$ when
$\alpha^{-1} =$ 1 $\AA$. For the case $\alpha^{-1} =$ 2 $\AA$ cf.
Fig.~\ref{fig-HT-beta9+11}(II). In other words, the strength of
the localization which determines the size of the formed polaron,
along with the density of states which characterizes the system,
are both two key factors for the conductivity and its dependance
on $F$ and $T$.

\subsection{Low Temperatures} \label{ResSubsec:LT}
\begin{figure}[]
\includegraphics[width=9cm]{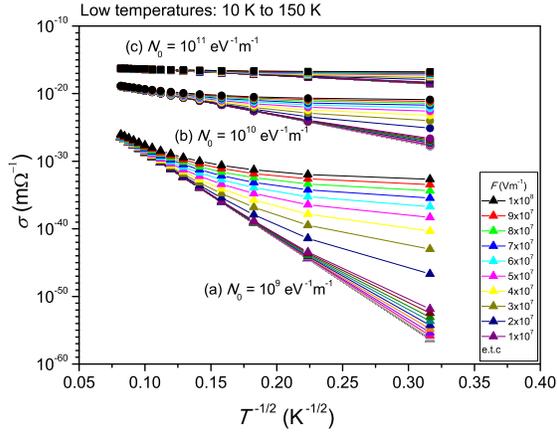}
\hspace*{-1cm} %\vspace*{-0.7cm}
\caption{\label{fig-LT-T12}
$\sigma$ versus $T^{-1/2}$ for cases
(a) $N_{0}=10^{9}$ eV$^{-1}$m$^{-1}$,
(b) $N_{0}=10^{10}$ eV$^{-1}$m$^{-1}$ and
(c) $N_{0}=10^{11}$ eV$^{-1}$m$^{-1}$.
$\alpha^{-1}=2$ $\AA$.
$T=(10$ - $150)$ K and
$F=(5\times 10^{3}$ - $1\times 10^{8})$ Vm$^{-1}$.}
\end{figure}

For the temperature range $T=(10$ - $150)$ K,
Fig.~\ref{fig-LT-T12} presents $\sigma$ as a function of $T^{-1/2}$
for different magnitudes of the DOS, i.e.
$N_{0}=(10^{9}$ - $10^{11})$ eV$^{-1}$m$^{-1}$ [cases (a) to (c), respectively].
$\alpha^{-1}=2$ $\AA$.
$F=(5\times 10^{3}$ - $1\times 10^{8})$ Vm$^{-1}$.
Again, we depict our results as a function of $T^{-1/2}$ in order
to compare them with the analytically obtained formula
$\ln\sigma\propto T^{-1/2}$ which holds for low up to moderate
electric fields and was previously obtained by two of us
\cite{DT:2010:Fcr} following a different theoretical treatment and
taking into account the effect of correlations, namely
\begin{equation}
\sigma^{l,cr}(F,T)\propto
\exp\left[-\left(\frac{T_{0}^{l,cr}}{T}\right)^{1/2} \!\!\!\!\!\! \left(1-\frac{F^{2}}{g(T)}\right)^{1/2}\right].
\end{equation}
Here $T_{0}^{l,cr}=1.96\alpha/k_{B}N_{0}$.
We observe that higher density of states leads to
higher conductivity
in such a way that $\frac{d\sigma}{dN_0}$ is smaller
for higher densities of states.
In total, in this specific example,
augmenting DOS by two orders of magnitude
increases the conductivity by approximately
tens of orders of magnitude.
From these results we realize that the $T^{-1/2}$-behaviour of $\sigma$
\cite{DT:2010:Fcr} holds for low up to moderate electric fields.
For higher values of $F$ the conductivity deviates from
the $T^{-1/2}$-behaviour as $T$ decreases and
this deviation appears to be larger the stronger the electric field is.
This deviation is due to the constructive energetic contribution of the
electric field which leads to the increase of the number of
available sites that can host the carrier, i.e. the range
$\overline{\Re}_{nn}^{l}$ does not depend solely on $T$. As the
temperature further decreases, for strong enough electric fields,
the range $\overline{\Re}_{nn}^{l}$ depends exclusively on the
applied electric field, as essentially all hops are downward in
energy. As a result, the conductivity does not depend on the
temperature. In other words,
there is  a transition from thermally-assisted to field-assisted hopping.

We remind the reader that in the \emph{high temperature} range
$T=(160$ - $300)$ K discussed in subsection ~\ref{ResSubsec:HT},
the electric field affects the range between two sites in the
``hopping space'' both for the absorption and the emission branch,
leading also to the appearance of the destructive role of the
electric field. In contrast, this effect does not appear here in
the \emph{low temperature} range $T=(10$ - $150)$ K, because in
the corresponding expression for the range between two sites in
the ``hopping space'' the electric field affects only the finite area
of the absorption branch.

\begin{figure*}[]
\includegraphics[width=10cm]{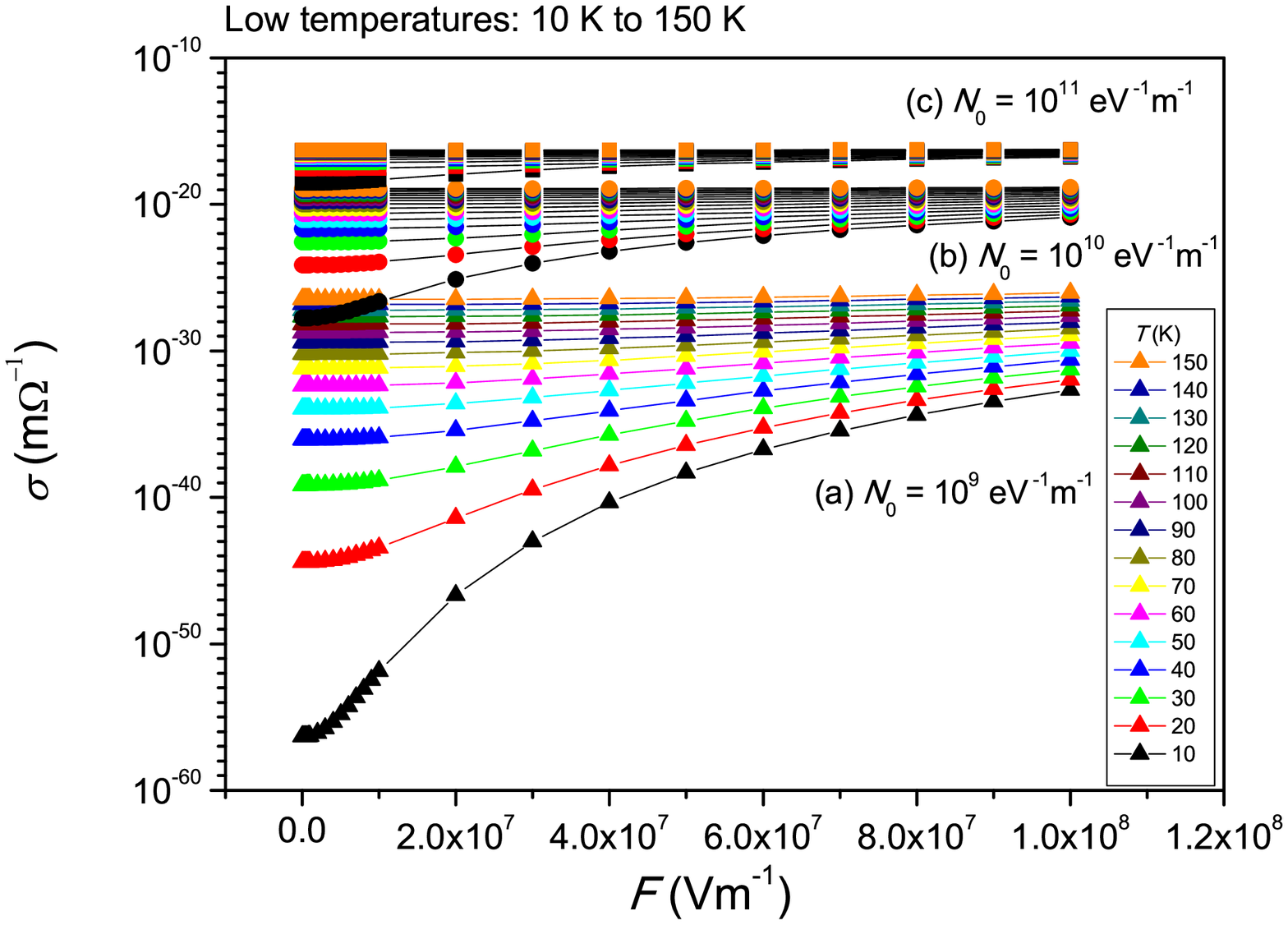}
\hspace*{-1cm} %\vspace*{-0.5cm}
\includegraphics[width=8cm]{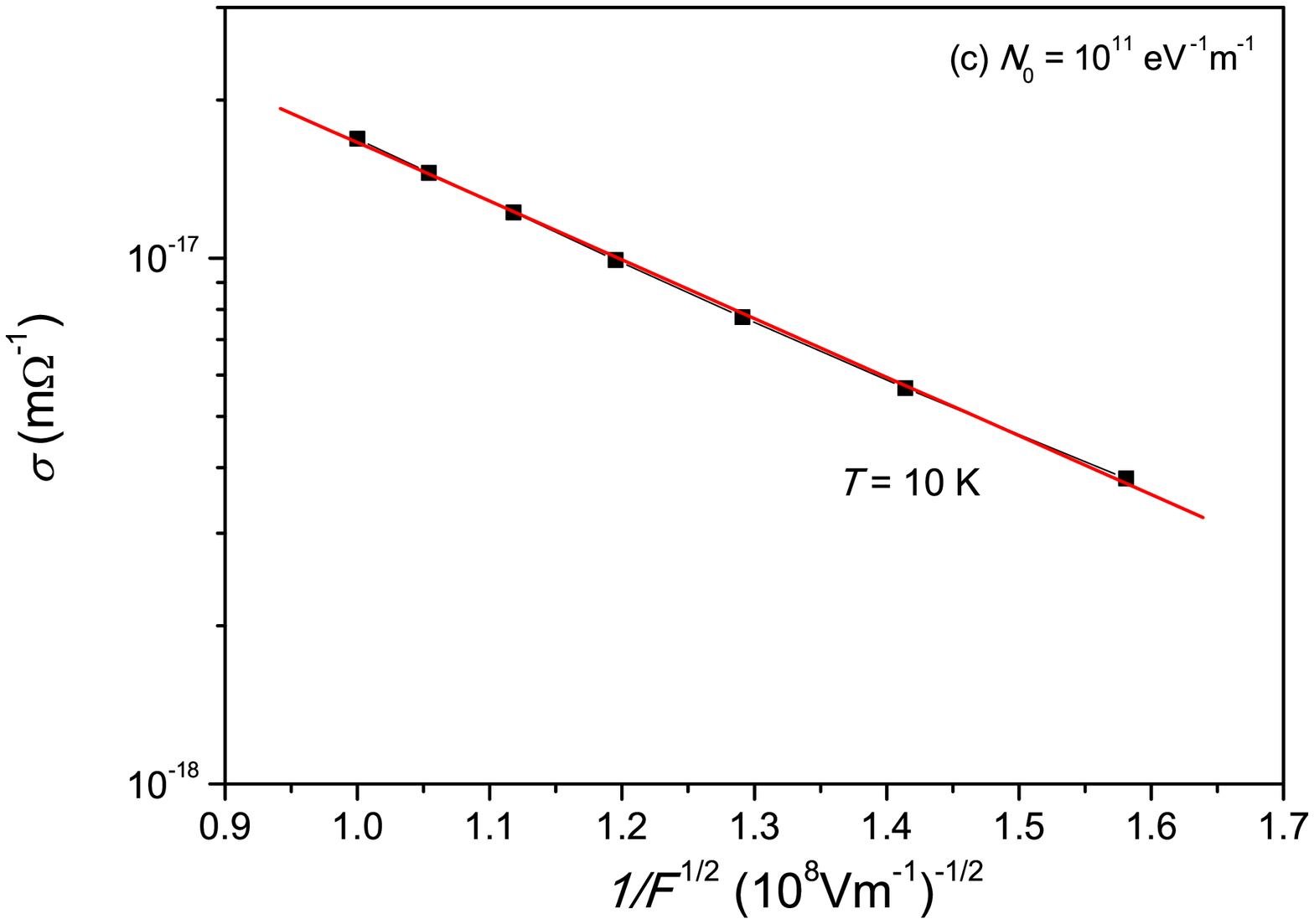}
\caption{\label{fig-LT-Fa-Fc}
(I) Left Panel.
$\sigma$ versus $F$ varying the density of states, i.e. for cases
(a) $N_{0}=10^{9}$ eV$^{-1}$m$^{-1}$,
(b) $N_{0}=10^{10}$ eV$^{-1}$m$^{-1}$ and
(c) $N_{0}=10^{11}$ eV$^{-1}$m$^{-1}$.
$F=(5\times 10^{3}$ - $1\times10^{8})$ Vm$^{-1}$ and
$T=(10$ - $150)$ K.
(II) Right Panel.
$\sigma$ versus $F^{-1/2}$ for case
(c) $N_{0}=10^{11}$ eV$^{-1}$m$^{-1}$ at $T=10$ K and
$F=(4\times 10^{7}$ - $1\times 10^{8})$ Vm$^{-1}$.
In both panels $\alpha^{-1}=2$ $\AA$.}
\end{figure*}

Figure \ref{fig-LT-Fa-Fc}(I) presents the conductivity for
different densities of states
$N_{0}=10^{9}$ - $10^{11}$ eV$^{-1}$m$^{-1}$ as a function of the applied electric field.
For low up to moderate electric fields the conductivity follows nicely
the $F^{2}$-behaviour, as we expected from the analytical
expression previously reported \cite{TD:2009:F,DT:2010:Fcr}.
Specifically, when the condition $F^{2}/g(T)\ll1$ is satisfied, i.e.
$e\alpha^{-1}F\ll2k_{B}T$, two of us have showed \cite{DT:2010:Fcr}
\begin{equation}\label{conduct-F-LT-cr}
\ln\sigma^{l,cr}(F,T)\propto\ln\sigma^{l,cr}(0,T)+h(F)/f^{l,cr}(T),
\end{equation}
where $\ln\sigma^{l,cr}(0,T)=-(T_{0}^{l,cr}/T)^{1/2}$, $h(F)=F^{2}$ and
$f^{l,cr}(T)=[(T_{0}^{l,cr}/T)^{1/2}/2g(T)]^{-1}$.
Increasing the electric field the conductivity
becomes independent of the temperature and follows a
$1/F^{1/2}$-behaviour (Fig.~\ref{fig-LT-Fa-Fc}(II)). The linear fit
is of the form $y = A + B x$ with $A = -15.664 \pm 0.018$, $B = -
11159.837 \pm 142.759$ and $R = -0.99959 $. In the region between
the $F^{2}$ and $1/F^{1/2}$-behaviour, $\ln \sigma$ increases
almost linearly with $F$.

\begin{figure*}[]
\includegraphics[width=9cm]{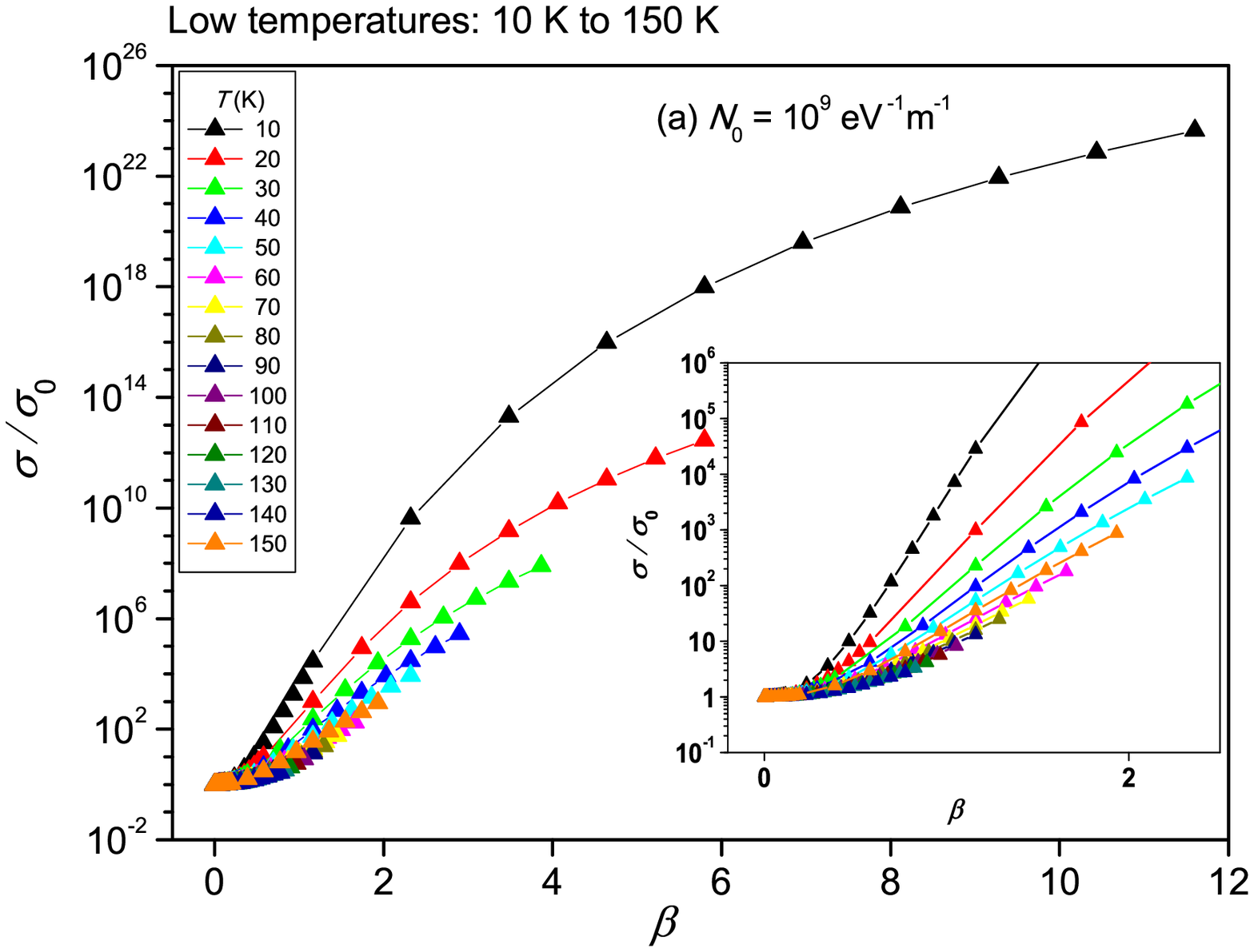}
\hspace*{-1cm}
\includegraphics[width=9cm]{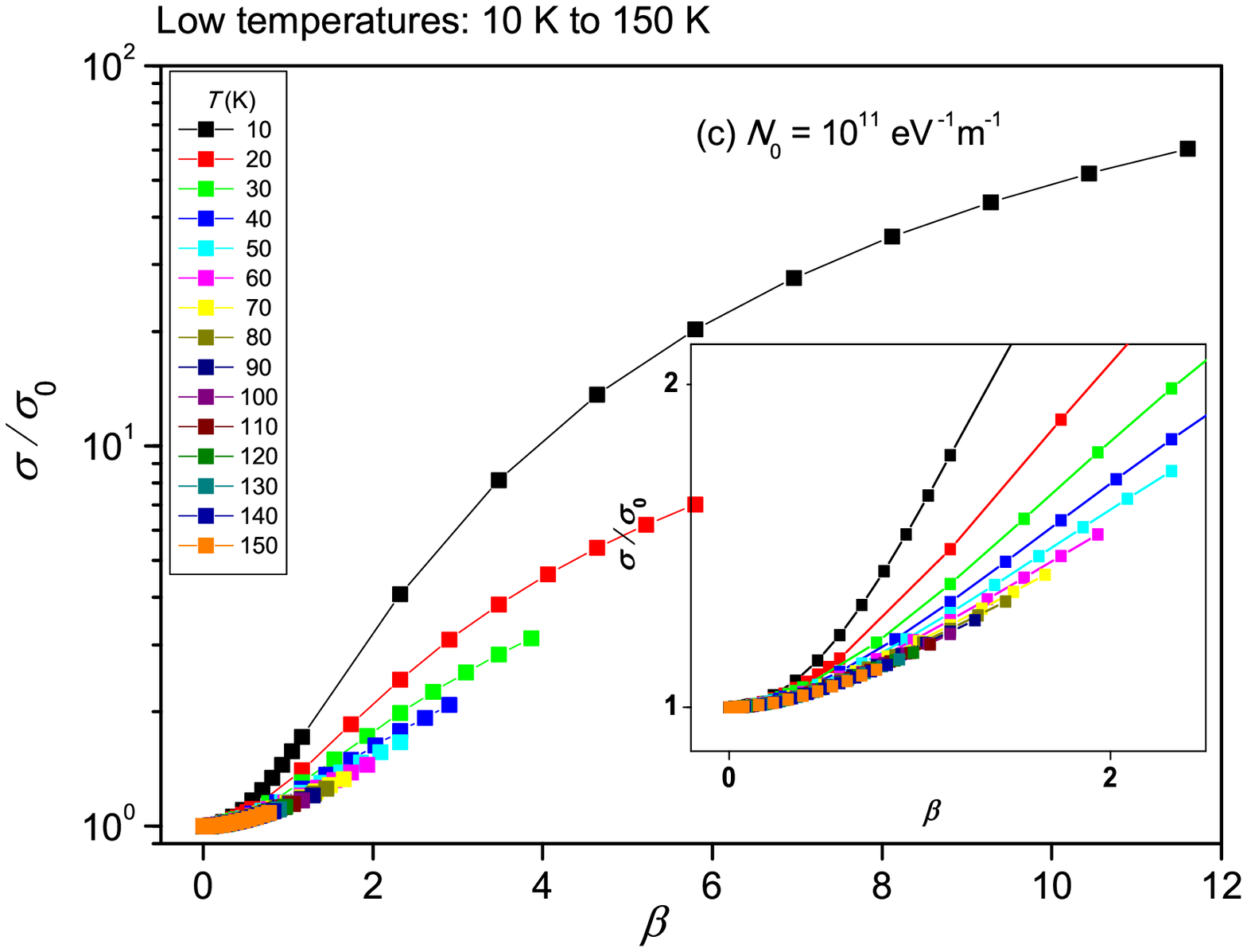}
\vspace*{-0.5cm}
\caption{\label{fig-LT-beta9+11}
$\sigma/\sigma_{0}$ versus $\beta = eF/2\alpha k_{B}T$ for cases (a)
$N_{0}=10^{9}$ eV$^{-1}$m$^{-1}$ and (c) $N_{0}=10^{11}$
eV$^{-1}$m$^{-1}$. $\alpha^{-1}=2$ $\AA$.
$F=(5\times 10^{3}$ - $1\times 10^{8})$
Vm$^{-1}$ and $T=(10$ - $150)$ K. The insets present enlargements
of $\sigma/\sigma_{0}$ versus $eF/2\alpha k_{B}T$.}
\end{figure*}

The influence of both $F$ and $T$ on $\sigma$ is shown in
Fig.~\ref{fig-LT-beta9+11}, where we depict $\sigma/\sigma_{0}$
versus $\beta = eF/2\alpha k_{B}T$, for (a) $N_{0}=10^{9}$
eV$^{-1}$m$^{-1}$ and (c) $N_{0}=10^{11}$ eV$^{-1}$m$^{-1}$.
$F=(5\times 10^{3}$ - $1\times 10^{8})$ Vm$^{-1}$ and $T=(10$ -
$150)$ K. The influence of $F$ on $\sigma$ depends on $T$, and it
is greater at lower temperatures, while at higher temperatures the
influence of $F$ decreases significantly. Comparing
Fig.~\ref{fig-LT-beta9+11} with the corresponding one
for \emph{high temperatures} (cf. Fig.~\ref{fig-HT-beta9+11})
we observe that the dependence of $\sigma/\sigma_{0}$ on $\beta = eF/2\alpha k_{B}T$
for \emph{low temperatures} is very strong especially in contrast to the corresponding
variation at \emph{high temperatures} studied earlier on
Subsection~\ref{ResSubsec:HT}.

Fogler and Kelley~\cite{FoglerKelley:2005}, Raikh and Ruzin
\cite{RaikhRuzin:1989}, and Ma \emph{et al.} \cite{Ma:2007} refer
also to the transition of a 1D disordered \emph{electron} system
from the ohmic to the non-ohmic behaviour, and in this respect
their results are consistent with our results, for both \emph{low}
and \emph{high temperatures}. For strong enough electric fields,
Fogler and Kelley~\cite{FoglerKelley:2005}, Pollak and
Riess~\cite{PollakRiess:1976}, and
Shklovskii~\cite{Shklovskii:1973}, claim that in strong electric
fields only the forward hops need to be considered, in contrast to
Apsley and Hughes~\cite{ApsleyHughes:1975} who integrate over the
entire space. Our methodology follows in that sense Apsley and
Hughes~\cite{ApsleyHughes:1975} summing over forward as well as
backward hops in our 1D polaron system. Bourbie \emph{et
al.}~\cite{Bourbie:2004} study the $F$-dependence of the hopping
conductivity in disordered \emph{3D electron} systems. They
propose --among other mechanisms-- that $\sigma$ decreases with
increasing $F$, when $F$ is strong enough to affect the tunneling
probability. This is due to the influence of $F$ on the number of
percolation paths, in the sense that increasing $F$ certain paths
become disallowed. This is in analogy with our discussion about
the destructive role of $F$ at high $F$ and low DOS. Bourbie
\emph{et al.} \cite{Bourbie:2007}, taking into account that ``$F$
affects the effective dimension of the transport path, reducing it
in the high-$F$ regime to 1D'', showed that when $F = 10^{8}$
Vm$^{-1}$, $T \approx$ (200 - 330) K, $\alpha^{-1} =$ 2 $\AA$ and
$N_{0}=10^{12}$ eV$^{-1}$m$^{-1}$, the conductivity decreases with
increasing temperature. This has been attributed to the
competition between thermal-assisted and field-assisted hopping.
We have obtained a similar behaviour for $\sigma$ when $F =
10^{8}$ Vm$^{-1}$, at \emph{high temperatures} $T =$ (160-300) K,
$\alpha^{-1} =$ 2$\AA$ and $N_{0}=10^{13}$ eV$^{-1}$m$^{-1}$ (cf.
Fig.~\ref{fig-HT-ALLDOS}). Bourbie \emph{et al.} have also
included different forms of the DOS and mention that these
different DOS lead to very similar $F$-dependence of $\sigma$. D.
Bourbie \cite{Bourbie:2011} also used some different values for
the extent of the \emph{3D electronic} wave function arriving at
the result that greater extent of the carrier leads to higher
conductivity in analogy with our results for \emph{1D polarons}.
However, we underline that all the above works
\cite{FoglerKelley:2005,ApsleyHughes:1975,PollakRiess:1976,Shklovskii:1973,Bourbie:2004,Bourbie:2007,Bourbie:2011}
refer to \emph{electrons} while we study \emph{polarons}.
Moreover, in our work we have scrutinized the importance of the
magnitude of the density of states and the spatial extent of the
localized electronic wave function (for arbitrary electric fields
up to the polaron dissociation limit and for any ``reasonable'' temperature).
Finally, in 1D systems the ionic or protonic transport might play a role in some cases
\cite{JiaxiongWang:2008,Pavlenko:2000,Pavlenko:2003,Rak:2006}.
However, in the present manuscript we do not investigate such possibilities.

\section{Conclusion}\label{Sec:conclusion}
We showed that the strength of the localization which determines the size of the formed polaron
along with the density of states are two key factors for the conductivity and its dependence on
the electric field and the temperature either at \emph{high} or at \emph{low temperatures}.
These aspects of small polaron hopping have been nearly ignored in the past.

To accomplish our task, we developed a novel theoretical approach
inspired by the eminent work of Apsley and
Hughes~\cite{ApsleyHughes:1975} in combination with the
GMCM~\cite{TSK:2005,TD:2009,TDRev:2009,TD:2009:F,DT:2010:Fcr} and
references therein. In addition, we combined analytical work with
numerical calculations. In the present model the expression which
determines the conductivity (cf. Eq.~\ref{condu}) depends on both
the density of states and the extent of the electronic wave
function. We varied the DOS by few orders of magnitude near values
which are relevant to common 1D
systems~\cite{HKS:2010,HKS:2011:erratum,Bourbie:2007,Bourbie:2011}
and the extent of the electronic wave function from 1 to 5 $\AA$,
i.e. for reasonable values for common organic
molecules~\cite{HKS:2009,HSK:2009}. Although in the present
manuscript we used for simplicity a constant density of states, it
is evident from Eq.~\ref{condu} that one could also try an energy
dependent DOS via the same approach. We examined $\sigma(T,F)$ for
temperatures from 10 up to 300 K and up to the electric field
values where polarons dissociate ($\approx 1 \times 10^8$
Vm$^{-1}$).

We showed the the electric field plays both a \emph{constructive} role by offering energy for the polaron hops and
a \emph{destructive} one,
in the sense that the stronger it is the more it forces the polaron to jump opposite to
the $\bm{F}$ direction prohibiting forward jumps to neighboring sites.
The relative strength of these two roles depends on the DOS and localization regimes.

Our present method confirms that either for \emph{high temperatures} or for \emph{low temperatures},
higher density of states leads to higher conductivity.
This is done in such a way that $\frac{d\sigma}{dN_0}$ is smaller for higher densities of states.
Conclusively, augmenting DOS by few orders of magnitude increases the conductivity by many orders of magnitude.

For \emph{high temperatures}, for the smaller densities of states
$\frac{d\sigma}{dF} < 0$ i.e. the conductivity is larger for lower $F$.
This is due to the competitive role of the directionality imposed by the electric field and
the temperature. This directionality affects destructively $\sigma$
when only few sites are available for the polaron i.e. for small DOS.
We noticed that according to Eq.~\ref{rangeHT-hopping} and Fig.~\ref{fig-Drawing1+2}(I),
the electric field affects the range between two sites in the ``hopping
space'' both for the absorption and the emission branch.
On the contrary, this effect does not appear at \emph{low temperatures},
because in the corresponding expression for the range between two sites in
the ``hopping space'' at \emph{low temperatures} (Eq.~\ref{rangeLT-hopping} and Fig.~\ref{fig-Drawing3+4}(I)),
the electric field affects only the finite area of the absorption branch.
We also noticed that the electric field plays a constructive role, too,
due to its energy offer to the polarons.
For ``medium'' DOS the available sites are numerous enough so that the directionality of $F$
hardly affects the conductivity.
For higher DOS only the constructive energetic influence of
the electric field appears. Now $\frac{d\sigma}{dF} > 0$ i.e. the
conductivity is larger for higher $F$.
Finally, we observed that the temperature has a greater effect on
the conductivity, the smaller the density of states is.

Our results confirmed that either for \emph{high} or for \emph{low temperatures}
the behaviour $\ln\sigma\propto T^{-1/2}$
previously obtained by two of us~\cite{DT:2010:Fcr},
following a different theoretical treatment and taking into account the effect of correlations,
holds for low up to moderate electric fields.
Moreover, for low electric fields
the conductivity follows the $F^{2}$-behaviour~\cite{TD:2009:F,DT:2010:Fcr},
and increasing $F$ the conductivity becomes independent of $T$ and
it follows a $1/F^{1/2}$-behaviour while in the region between
the $F^{2}$ and $1/F^{1/2}$-behaviour, $\ln \sigma$ increases almost linearly with $F$.

We examined the deviation of conductivity from its ohmic value
under the influence of both the external stimuli $F$ and $T$ (introducing $\beta = eF/2\alpha k_{B}T$).
This was done either for \emph{high} or for \emph{low temperatures},
and for different DOS. We showed that $\sigma(\beta)$ depends strongly on the value of the DOS,
and either decreasing or increasing $\sigma(\beta)$ could be observed.
We noticed that the variation of $\sigma/\sigma_{0}$ versus $\beta$
is generally very small in \emph{high temperatures} compared to
the corresponding variation at \emph{low temperatures}.

Finally, we studied the conductivity for different values of the
spatial extent of the localized electronic wave function in the
range $\alpha^{-1}=(1$ - $5)$ $\AA$. Our results confirm that more
localized polarons exhibit smaller conductivity. Particularly,
five times increase of $\alpha^{-1}$ lead to two orders of
magnitude greater conductivity. Moreover, we showed that
$\frac{d\sigma}{d\beta}>0$ for any $T$ when $\alpha^{-1} =$ 3, 4
and 5 $\AA$, while $\frac{d\sigma}{d\beta} < 0$ when $\alpha^{-1}
=$ 1 $\AA$. For the case $\alpha^{-1} =$ 2 $\AA$ we observed an
intermediate behaviour.

In summary, we proved that the size of the polaron
and the density of states are crucial factors for the behaviour of the conductivity
and  its dependence on the electric field and the temperature
either at \emph{high} or at \emph{low temperatures}.\\

{\bf Acknowledgments}
C. S. acknowledges ELKE (National and Kapodistrian University of Athens) for financial support.

%\begin{widetext}

\section*{Appendix}
The integrals $I_{1}$ and $I_{2}$, relevant at \emph{high
temperatures}, are given below:

\begin{equation}\hspace{-2cm}
I_{1}=\sum_{0,\pi}\int_{\frac{1}{3}(E'_{i}-\beta\overline{\Re}_{nn}^{h}\cos\theta)}^{\frac{1}{3}(2\overline{\Re}_{nn}^{h}+E'_{i})}
N(E'_{j})[1-f(E'_{j})]
\left(\frac{\overline{\Re}_{nn}^{h}-\frac{3}{2}E'_{j}+\frac{1}{2}E'_{i}}{1+\frac{\beta}{2}\cos\theta}\right)\cos\theta
dE'_{j}.
\end{equation}
\begin{equation}\hspace{-2cm}
I_{2}=\sum_{0,\pi}\int_{\frac{1}{3}(E'_{i}-\beta\overline{\Re}_{nn}^{h}\cos\theta)}^{\frac{1}{3}(2\overline{\Re}_{nn}^{h}+E'_{i})}
N(E'_{j})[1-f(E'_{j})]dE'_{j}.
\end{equation}
As the integrals $I_{1}$ and $I_{2}$ diverge at
$\beta=-2/\cos\theta$, we change variables and integrate over
$R'$. Hence:
\begin{eqnarray}\hspace{-2cm}
I_{1}=\sum_{0,\pi}\int_{0}^{\overline{\Re}_{nn}^{h}}
N\left(\frac{2}{3}\left[\overline{\Re}_{nn}^{h}-R'\left(1+\frac{\beta}{2}\cos\theta\right)+\frac{E'_{i}}{2}\right]\right)  \times \nonumber \\
\left[1-f\left(\frac{2}{3}\left[\overline{\Re}_{nn}^{h}-R'\left(1+\frac{\beta}{2}\cos\theta\right)+\frac{E'_{i}}{2}\right]\right)\right]R'\cos\theta
dR'.
\end{eqnarray}
\begin{eqnarray}\hspace{-2cm}
I_{2}=\sum_{0,\pi}\int_{0}^{\overline{\Re}_{nn}^{h}}
N\left(\frac{2}{3}\left[\overline{\Re}_{nn}^{h}-R'\left(1+\frac{\beta}{2}\cos\theta\right)+\frac{E'_{i}}{2}\right]\right)  \times \nonumber \\
\left[1-f\left(\frac{2}{3}\left[\overline{\Re}^{h}_{nn}-R'\left(1+\frac{\beta}{2}\cos\theta\right)+\frac{E'_{i}}{2}\right]\right)\right]dR'.
\end{eqnarray}

The integrals $I_{1}$, $I_{2}$, $I_{3}$, $I_{4}$ relevant at
\emph{low temperatures} are given below.
\begin{equation}\hspace{-2cm}
I_{1}=\sum_{0,\pi}\int_{E^{*}_{i}-\beta\overline{\Re}_{nn}^{l}\cos\theta}^{E^{*}_{i}+\overline{\Re}_{nn}^{l}}
N(E^{*}_{j})[1-f(E^{*}_{j})]\left(\frac{\overline{\Re}_{nn}^{l}-E^{*}_{j}+E^{*}_{i}}{1+\beta\cos\theta}\right)\cos\theta
dE^{*}_{j}.
\end{equation}
\begin{equation}\hspace{-2cm}
I_{2}=\sum_{0,\pi}\int_{-\infty}^{E^{*}_{i}-\beta\overline{\Re}_{nn}^{l}\cos\theta}N(E^{*}_{j})
[1-f(E^{*}_{j})]\overline{\Re}_{nn}^{l}\cos\theta dE^{*}_{j}.
\end{equation}
\begin{equation}\hspace{-2cm}
I_{3}=\sum_{0,\pi}\int_{E^{*}_{i}-\beta\overline{\Re}_{nn}^{l}\cos\theta}^{E^{*}_{i}+\overline{\Re}_{nn}^{l}}N(E^{*}_{j})
[1-f(E^{*}_{j})]dE^{*}_{j}.
\end{equation}
\begin{equation}\hspace{-2cm}
I_{4}=\sum_{0,\pi}\int_{-\infty}^{E^{*}_{i}-\beta\overline{\Re}_{nn}^{l}\cos\theta}
N(E^{*}_{j})[1-f(E^{*}_{j})]dE^{*}_{j}.
\end{equation}
As the integrals $I_{1}$ and $I_{3}$ diverge at
$\beta=-2/\cos\theta$, we change variables and integrate over
$R'$. Hence:
\begin{equation}\hspace{-2cm}
I_{1}=\sum_{0,\pi}\int_{0}^{\overline{\Re}_{nn}^{l}}
N(\overline{\Re}_{nn}^{l}-R'(1+\beta\cos\theta)+E^{*}_{i}))[1-f(\overline{\Re}_{nn}^{l}-R'(1+\beta\cos\theta)+E^{*}_{i})]R'\cos\theta
dR'.
\end{equation}
\begin{equation}\hspace{-2cm}
I_{3}=\sum_{0,\pi}\int_{0}^{\overline{\Re}_{nn}^{l}}
N(\overline{\Re}_{nn}^{l}-R'(1+\beta\cos\theta)+E^{*}_{i}))[1-f(\overline{\Re}_{nn}^{l}-R'(1+\beta\cos\theta)+E^{*}_{i})]dR'.
\end{equation}

%\end{widetext}

\end{document}